\documentclass[fleqn]{article}
\usepackage{longtable}
\usepackage{graphicx}

\begin {document}
\begin{flushleft}
{\LARGE
{\bf Energy Levels and Radiative Rates for Transitions in F-like Sc~XIII and Ne-like Sc~XII and Y~XXX}
}\\

\vspace{1.5 cm}

{\bf K.  M.  ~Aggarwal} \\ 

\vspace*{1.0cm}

Astrophysics Research Centre, School of Mathematics and Physics, Queen's University Belfast, Belfast BT7 1NN, Northern Ireland, UK\\ 

\vspace*{0.5 cm} 

e-mail: K.Aggarwal@qub.ac.uk \\

\vspace*{1.50cm}

Received  9 April 2018\\
Accepted for publication 25 April 2018 \\
Published xx  Month 2018 \\

\vspace*{1.5cm}

PACS Ref:  31.25 Jf, 32.70 Cs

\vspace*{1.0 cm}

\hrule

\end{flushleft}

\clearpage


\begin{abstract}

Energy levels, radiative rates and lifetimes  are reported for F-like Sc~XIII and Ne-like Sc~XII and Y~XXX for which the general-purpose relativistic atomic structure package ({\sc grasp}) has been adopted. For all three ions limited data exist in the literature but comparisons have been  made wherever possible to assess the accuracy of the calculations. In the present work the lowest 102, 125 and 139 levels have been considered for the respective ions. Additionally, calculations have also been performed with the flexible atomic code ({\sc fac}) to (particularly) confirm the accuracy of energy levels.
\end{abstract}

{\bf Keywords:} rare earth elements, energy levels,   radiative rates, lifetimes, F-like and Ne-like ions









\setcounter{section}{0} 

\section{Introduction}

Atomic data for several parameters, including energy levels and radiative rates, are required for the diagnostics and modelling of plasmas. Often the required data are not available experimentally and therefore it becomes necessary to obtain theoretical results. Consequently, a vast amount of theoretical data are available in the literature for a (very) wide range of ions. However, comparatively neglected are the rare earth elements (Sc, Y and those with 57 $\le$ Z $\le$ 71), although work on their ions is gradually picking momentum. In this paper we report atomic data for F-like Sc~XIII and Ne-like Sc~XII and Y~XXX.

The importance of Sc ions was realised early by Pryce \cite{mhlp} who speculated that their emission lines may be observed in the visible or near-visible regions of the coronal plasma. However, to the best of our knowledge no lines of Sc~XII or Sc~XIII have so far been observed in the astrophysical plasmas. This may also be confirmed by the {\sc
chianti}  ({\tt http://www.chiantidatabase.org/}) database, which stores data for all ions of astrophysical importance, but none for the Sc ions. Nevertheless, several lines of Sc~XIII have been measured by Jup\'{e}n et al. \cite{cj} in the laser-produced plasmas.  Similarly, energy levels and radiative rates ($A$-values) have been determined for several ions of the F sequence of which the work by J\"{o}nsson et al. \cite{jag} is not only the latest but also the most accurate, because differences with the measurements are minimal for the energy levels. Unfortunately, they calculated data only for the lowest three levels of the 2s$^2$2p$^5$ and 2s2p$^6$ configurations, which are not sufficient for modelling of plasmas, because the data are required for a wider range of levels. Therefore, in this work we cover a much wider range of levels, discussed in Section~2. We also note here that similar data for F-like Y~XXXI are not considered here, because results have already been reported in a separate paper \cite{flike1}. 

Ne-like ions are of interest for the studies of astrophysical, lasing and fusion plasmas, and therefore  many workers, such as \cite{cl, hag, zs2, pq, ah, jon}, have reported data for a wide range of ions. However, for brevity many of them have not reported data for Sc~XII, although Hibbert et al. \cite{ah} have listed lifetimes ($\tau$) for the lowest 27 levels of the 2s$^2$2p$^6$ and 2s$^2$2p$^5$3$\ell$  configurations. The only other results available for comparisons are those of  Cogordan and Lunell \cite{cl} and J\"{o}nsson et al. \cite{jon}, but are limited to the lowest 27 levels of the 2s$^2$2p$^6$ and 2s$^2$2p$^5$3$\ell$  configurations. Therefore, there is a clear need to expand the range of levels for this ion. Similarly, limited results are available in the literature for Y~XXX, mainly by \cite{cl, hag, zs2, pq}, whereas Nilsen and Scofield \cite{ns} and Silwal et al. \cite{yr} have measured wavelengths for a few transitions of this ion, which is of particular interest for the diagnostics of tokamak fusion plasmas, as it is one of the impurity elements.  Therefore, our {\em aim} is to report a complete set of data for energies and lifetimes for a larger number of levels for all three above named ions and $A$-values for all transitions among their levels, not only for the dominant allowed electric dipole (E1) type, but also for electric quadrupole (E2), magnetic dipole (M1), and magnetic quadrupole (M2),  which are not only required for complete and reliable  plasma models,  but are also useful for more accurate determination of lifetimes.

\section{Energy levels} 

As in our earlier work on several F-like \cite{flike1, flike2} and Ne-like \cite{nelike1, nelike2} ions, we adopt the fully relativistic {\sc grasp} (general-purpose relativistic atomic structure package) code of Grant et al. \cite{grasp0} to determine the atomic structure, and subsequently to calculate energy levels and $A$-values. However, this earlier version has been significantly revised by Dr. P.H. Norrington (one of the authors), and is currently hosted at the website: {\tt http://amdpp.phys.strath.ac.uk/UK\_APAP/codes.html}. Similarly,  the option of {\em extended average level} (EAL), in which a weighted (proportional to 2$j$+1) trace of the Hamiltonian matrix is minimized, is used.  This
produces a compromise set of orbitals describing closely-lying states with  moderate accuracy. This also provides comparable results with other options such as average level (AL). 

Since both Sc and Y are moderately heavy elements, both relativistic effects and configuration interaction (CI) are important for the determination of atomic structures. Our adopted version of the {\sc grasp} code is fully relativistic, as are the other ones, such as {\sc grasp2k} \cite{grasp2k}, but it cannot handle the inclusion of an extensive  CI, or a very large number of configuration state functions (CSF). Therefore, we have also performed calculations  with the  {\em Flexible Atomic Code} ({\sc fac}) of Gu \cite{fac}, hosted at the website {\tt https://www-amdis.iaea.org/FAC/}. This is also a fully relativistic code and provides a variety of atomic parameters.  Not only the code yields data which in most instances are  comparable to those generated with {\sc grasp}, the inclusion of a very large CI is also possible with ease and efficiency. Therefore, these parallel calculations serve two purposes, i.e. firstly, the accuracy of the determined energy levels can be assessed, and this is necessary because similar results for a majority of levels are not available with which to compare, as already stated in Section~1, and secondly, the effect of larger CI (if any) can be quantified.  

\subsection{Sc~XIII}

With {\sc grasp} we have performed a series of calculations with increasing CI, but mention here only three, namely (i) GRASP1,  which includes 113 levels of the 2s$^2$2p$^5$, 2s2p$^6$, 2s$^2$2p$^4$3$\ell$,  2s2p$^5$3$\ell$,  and 2p$^6$3$\ell$ (11) configurations, (ii) GRASP2, which includes further 8 of 2s$^2$2p$^4$4$\ell$ and  2s2p$^5$4$\ell$ giving rise to additional 159 levels,  and finally (iii) GRASP3, which includes 501 levels in total from 38  configurations, the additional  ones being 2p$^6$4$\ell$, 2s$^2$2p$^4$5$\ell$, 2s2p$^5$5$\ell$,  and 2p$^6$5$\ell$. However, for brevity we will discuss results from only our final calculations, but the effect of additional CI will be discussed with those from {\sc fac}.

\begin{table}
\caption{Energies (in Ryd) and lifetimes ($\tau$, s) for the lowest 102 levels of  Sc XIII. $a{\pm}b \equiv$ $a\times$10$^{{\pm}b}$.}
\small 
\centering
\begin{tabular}{rllrrrrrrrr} \hline
Index  &     Configuration        & Level                & NIST  &  HFR &  GRASP1 & GRASP2        &    FAC1 & FAC2 & FAC3     &    $\tau$   (GRASP2)   \\
 \hline		
    1  &  2s$^2$2p$^5$  	    &  $^2$P$^o_{3/2}$   &  0.000   &  0.0000	&  0.0000 &   0.0000   &  0.0000  &   0.0000   &  0.0000  &   ........ \\
    2  &  2s$^2$2p$^5$  	    &  $^2$P$^o_{1/2}$   &  0.345   &  0.3458	&  0.3543 &   0.3430   &  0.3426  &   0.3424   &  0.3425  &   1.045-03 \\
    3  &  2s2p$^6$		    &  $^2$S$  _{1/2}$   &  6.959   &  6.9588	&  7.1071 &   7.0978   &  7.1152  &   7.1021   &  7.0877  &   1.363-11 \\
    4  &  2s$^2$2p$^4$3s	    &  $^4$P$  _{5/2}$   & 32.016   & 32.0155	& 31.9299 &  31.9067   & 31.9416  &  31.9404   & 31.7915  &   9.843-11 \\
    5  &  2s$^2$2p$^4$3s	    &  $^4$P$  _{3/2}$   & 32.173   & 32.1703	& 32.0898 &  32.0666   & 32.1064  &  32.1073   & 31.9452  &   3.494-12 \\
    6  &  2s$^2$2p$^4$3s	    &  $^4$P$  _{1/2}$   & 32.324   & 32.3235	& 32.2361 &  32.2104   & 32.2440  &  32.2429   & 32.0925  &   6.617-11 \\
    7  &  2s$^2$2p$^4$3s	    &  $^2$P$  _{3/2}$   & 32.394   & 32.3937	& 32.3158 &  32.2906   & 32.3375  &  32.3419   & 32.1612  &   1.623-12 \\
    8  &  2s$^2$2p$^4$3s	    &  $^2$P$  _{1/2}$   & 32.570   & 32.5696	& 32.4940 &  32.4684   & 32.5216  &  32.5282   & 32.3353  &   1.093-12 \\
    9  &  2s$^2$2p$^4$3s	    &  $^2$D$  _{5/2}$   & 32.983   & 32.9833	& 32.9299 &  32.8984   & 32.9169  &  32.9188   & 32.7710  &   2.825-12 \\
   10  &  2s$^2$2p$^4$3s	    &  $^2$D$  _{3/2}$   & 32.995   & 32.9951	& 32.9402 &  32.9088   & 32.9285  &  32.9308   & 32.7803  &   2.613-12 \\
   11  &  2s$^2$2p$^4$3p	    &  $^4$P$^o_{5/2}$   & 	    & 33.5910	& 33.5112 &  33.4860   & 33.5231  &  33.5164   & 33.3813  &   4.776-10 \\
   12  &  2s$^2$2p$^4$3p	    &  $^4$P$^o_{3/2}$   & 	    & 33.6070	& 33.5246 &  33.5032   & 33.5404  &  33.5341   & 33.3978  &   4.386-10 \\
   13  &  2s$^2$2p$^4$3p	    &  $^4$P$^o_{1/2}$   & 	    & 33.7247	& 33.6439 &  33.6209   & 33.6564  &  33.6488   & 33.5135  &   4.412-10 \\
   14  &  2s$^2$2p$^4$3p	    &  $^4$D$^o_{7/2}$   & 	    & 33.7869	& 33.7148 &  33.6891   & 33.7260  &  33.7283   & 33.5782  &   2.996-10 \\
   15  &  2s$^2$2p$^4$($^3$P)3p     &  $^2$D$^o_{5/2}$   & 	    & 33.8297	& 33.7568 &  33.7331   & 33.7528  &  33.7554   & 33.6173  &   3.445-10 \\
   16  &  2s$^2$2p$^4$3s	    &  $^2$S$  _{1/2}$   & 33.885   & 33.8851	& 33.7722 &  33.7480   & 33.7750  &  33.7791   & 33.6554  &   2.766-12 \\
   17  &  2s$^2$2p$^4$($^3$P)3p     &  $^2$P$^o_{1/2}$   & 	    & 33.9692	& 33.9064 &  33.8823   & 33.9155  &  33.9094   & 33.7678  &   2.856-10 \\
   18  &  2s$^2$2p$^4$3p	    &  $^4$D$^o_{3/2}$   & 	    & 33.9882	& 33.9151 &  33.8892   & 33.9264  &  33.9289   & 33.7729  &   3.208-10 \\
   19  &  2s$^2$2p$^4$3p	    &  $^4$D$^o_{1/2}$   & 	    & 34.0260	& 33.9515 &  33.9265   & 33.9635  &  33.9658   & 33.8135  &   2.813-10 \\
   20  &  2s$^2$2p$^4$3p	    &  $^4$D$^o_{5/2}$   & 	    & 34.0802	& 34.0104 &  33.9827   & 34.0211  &  34.0248   & 33.8624  &   3.175-10 \\
   21  &  2s$^2$2p$^4$($^3$P)3p     &  $^2$P$^o_{3/2}$   & 	    & 34.1222	& 34.0295 &  34.0001   & 34.0359  &  34.0291   & 33.8819  &   2.730-10 \\
   22  &  2s$^2$2p$^4$3p	    &  $^4$S$^o_{3/2}$   & 	    & 34.1913	& 34.1203 &  34.0949   & 34.1350  &  34.1412   & 33.9804  &   1.730-10 \\
   23  &  2s$^2$2p$^4$($^3$P)3p     &  $^2$S$^o_{1/2}$   & 	    & 34.2408	& 34.1738 &  34.1455   & 34.1823  &  34.1861   & 34.0219  &   3.267-10 \\
   24  &  2s$^2$2p$^4$($^3$P)3p     &  $^2$D$^o_{3/2}$   & 	    & 34.2327	& 34.1795 &  34.1505   & 34.1919  &  34.1968   & 34.0268  &   2.310-10 \\
   25  &  2s$^2$2p$^4$3p	    &  $^2$F$^o_{5/2}$   & 	    & 34.5951	& 34.5491 &  34.5172   & 34.5357  &  34.5381   & 34.3973  &   3.788-10 \\
   26  &  2s$^2$2p$^4$3p	    &  $^2$F$^o_{7/2}$   & 	    & 34.6571	& 34.6149 &  34.5803   & 34.5979  &  34.5998   & 34.4607  &   3.335-10 \\
   27  &  2s$^2$2p$^4$($^1$D)3p     &  $^2$D$^o_{3/2}$   & 	    & 34.8134	& 34.7669 &  34.7375   & 34.7570  &  34.7635   & 34.6168  &   2.215-10 \\
   28  &  2s$^2$2p$^4$($^1$D)3p     &  $^2$D$^o_{5/2}$   & 	    & 34.8524	& 34.8089 &  34.7765   & 34.7965  &  34.8037   & 34.6550  &   2.420-10 \\
   29  &  2s$^2$2p$^4$($^1$D)3p     &  $^2$P$^o_{3/2}$   & 	    & 35.3175	& 35.2384 &  35.2135   & 35.2915  &  35.2895   & 35.0662  &   3.438-11 \\
   30  &  2s$^2$2p$^4$($^1$D)3p     &  $^2$P$^o_{1/2}$   & 	    & 35.3499	& 35.2733 &  35.2479   & 35.3197  &  35.2996   & 35.1242  &   3.100-11 \\
   31  &  2s$^2$2p$^4$($^1$S)3p     &  $^2$P$^o_{3/2}$   & 	    & 35.6572	& 35.5598 &  35.5353   & 35.5557  &  35.5546   & 35.4494  &   1.052-10 \\
   32  &  2s$^2$2p$^4$($^1$S)3p     &  $^2$P$^o_{1/2}$   & 	    & 35.7942	& 35.6876 &  35.6585   & 35.7274  &  35.7096   & 35.5420  &   1.080-10 \\
   33  &  2s$^2$2p$^4$3d	    &  $^4$D$  _{7/2}$   & 	    & 35.7683	& 35.6967 &  35.6671   & 35.7200  &  35.7015   & 35.5418  &   1.979-10 \\
   34  &  2s$^2$2p$^4$3d	    &  $^4$D$  _{5/2}$   & 	    & 35.7785	& 35.7053 &  35.6778   & 35.7450  &  35.7267   & 35.5494  &   1.877-10 \\
   35  &  2s$^2$2p$^4$3d	    &  $^4$D$  _{3/2}$   & 	    & 35.8200	& 35.7383 &  35.7131   & 35.7605  &  35.7436   & 35.5824  &   7.254-11 \\
   36  &  2s$^2$2p$^4$3d	    &  $^4$D$  _{1/2}$   & 	    & 35.8654	& 35.7854 &  35.7602   & 35.8086  &  35.7919   & 35.6295  &   6.519-11 \\
   37  &  2s$^2$2p$^4$3d	    &  $^4$F$  _{9/2}$   & 	    & 35.9488	& 35.8847 &  35.8552   & 35.8991  &  35.8891   & 35.7268  &   1.705-10 \\
   38  &  2s$^2$2p$^4$3d	    &  $^4$F$  _{7/2}$   & 	    & 36.0225	& 35.9661 &  35.9375   & 35.9733  &  35.9668   & 35.7922  &   1.598-10 \\
   39  &  2s$^2$2p$^4$3d	    &  $^4$P$  _{1/2}$   & 36.102   & 36.1017	& 36.0325 &  36.0065   & 36.0384  &  36.0374   & 35.8618  &   3.624-12 \\
   40  &  2s$^2$2p$^4$3d	    &  $^4$F$  _{5/2}$   & 36.162   & 36.1618	& 36.1014 &  36.0715   & 36.1102  &  36.1035   & 35.9298  &   2.505-12 \\
\hline
\end{tabular}
\end{table}

\clearpage
\newpage
\setcounter{table}{0} 
\begin{table}
\caption{... continued.}
\small 
\centering
\begin{tabular}{rllrrrrrrrrr} \hline
Index  &     Configuration        & Level                & NIST  &  HFR &  GRASP1 & GRASP2        &    FAC1 & FAC2 & FAC3     &    $\tau$   (GRASP2)   \\
 \hline 
   41  &  2s$^2$2p$^4$3d	    &  $^4$P$  _{3/2}$   & 36.162   & 36.1618	& 36.1030 &  36.0757   & 36.1108  &  36.1083   & 35.9327  &   2.022-12 \\
   42  &  2s$^2$2p$^4$($^3$P)3d     &  $^2$P$  _{1/2}$   & 36.215   & 36.2147	& 36.1592 &  36.1328   & 36.1694  &  36.1614   & 35.9864  &   2.918-12 \\
   43  &  2s$^2$2p$^4$3d	    &  $^4$F$  _{3/2}$   & 	    & 36.2171	& 36.1703 &  36.1415   & 36.1833  &  36.1737   & 36.0088  &   1.689-11 \\
   44  &  2s$^2$2p$^4$($^3$P)3d     &  $^2$F$  _{7/2}$   & 	    & 36.2271	& 36.1716 &  36.1404   & 36.1762  &  36.1690   & 35.9923  &   1.708-10 \\
   45  &  2s$^2$2p$^4$3d	    &  $^4$P$  _{5/2}$   & 36.257   & 36.2575	& 36.1973 &  36.1672   & 36.2024  &  36.1985   & 36.0251  &   3.312-12 \\
   46  &  2s$^2$2p$^4$($^3$P)3d     &  $^2$D$  _{3/2}$   & 36.307   & 36.3067	& 36.2530 &  36.2252   & 36.2604  &  36.2531   & 36.0767  &   7.388-13 \\
   47  &  2s$^2$2p$^4$($^3$P)3d     &  $^2$F$  _{5/2}$   & 36.336   & 36.3359	& 36.2848 &  36.2537   & 36.2843  &  36.2816   & 36.0964  &   4.289-12 \\
   48  &  2s$^2$2p$^4$($^3$P)3d     &  $^2$P$  _{3/2}$   & 36.453   & 36.4534	& 36.4051 &  36.3721   & 36.4022  &  36.3954   & 36.2186  &   1.771-12 \\
   49  &  2s$^2$2p$^4$($^3$P)3d     &  $^2$D$  _{5/2}$   & 36.494   & 36.5236	& 36.4578 &  36.4230   & 36.4572  &  36.4527   & 36.2662  &   4.378-13 \\
   50  &  2s$^2$2p$^4$3d	    &  $^2$G$  _{7/2}$   & 	    & 36.7662	& 36.7310 &  36.6935   & 36.7185  &  36.7056   & 36.5569  &   1.725-10 \\
   51  &  2s$^2$2p$^4$3d	    &  $^2$G$  _{9/2}$   & 	    & 36.7702	& 36.7362 &  36.6980   & 36.7210  &  36.7076   & 36.5605  &   1.855-10 \\
   52  &  2s$^2$2p$^4$($^1$D)3d     &  $^2$S$  _{1/2}$   & 36.945   & 36.9446	& 36.9226 &  36.8871   & 36.8926  &  36.8953   & 36.7196  &   1.683-13 \\
   53  &  2s$^2$2p$^4$($^1$D)3d     &  $^2$F$  _{5/2}$   & 36.971   & 36.9710	& 36.9441 &  36.9105   & 36.9207  &  36.9199   & 36.7583  &   3.243-12 \\
   54  &  2s$^2$2p$^4$($^1$D)3d     &  $^2$F$  _{7/2}$   & 	    & 37.0086	& 36.9805 &  36.9442   & 36.9522  &  36.9526   & 36.7894  &   1.494-10 \\
   55  &  2s$^2$2p$^4$($^1$D)3d     &  $^2$P$  _{3/2}$   & 37.103   & 37.1041	& 37.0808 &  37.0477   & 37.0676  &  37.0699   & 36.8950  &   1.329-13 \\
   56  &  2s$^2$2p$^4$($^1$D)3d     &  $^2$D$  _{5/2}$   & 37.104   & 37.1041	& 37.1013 &  37.0663   & 37.0875  &  37.0874   & 36.8952  &   1.644-13 \\
   57  &  2s$^2$2p$^4$($^1$D)3d     &  $^2$D$  _{3/2}$   & 37.219   & 37.1889	& 37.2125 &  37.1749   & 37.1913  &  37.1936   & 36.9972  &   1.723-13 \\
   58  &  2s$^2$2p$^4$($^1$D)3d     &  $^2$P$  _{1/2}$   & 	    & 37.2186	& 37.2138 &  37.1737   & 37.1914  &  37.1949   & 37.0132  &   1.250-13 \\
   59  &  2s$^2$2p$^4$($^1$S)3d     &  $^2$D$  _{5/2}$   & 37.817   & 37.8167	& 37.7296 &  37.7007   & 37.6980  &  37.6961   & 37.5840  &   1.283-12 \\
   60  &  2s$^2$2p$^4$($^1$S)3d     &  $^2$D$  _{3/2}$   & 37.873   & 37.8732	& 37.7955 &  37.7637   & 37.7723  &  37.7703   & 37.6458  &   3.332-13 \\
   61  &  2s2p$^5$3s		    &  $^4$P$^o_{5/2}$   & 	    & 	     	& 38.2026 &  38.1736   & 38.2253  &  38.2262   & 38.1003  &   2.172-11 \\
   62  &  2s2p$^5$3s		    &  $^4$P$^o_{3/2}$   & 	    & 	     	& 38.3683 &  38.3358   & 38.3890  &  38.3892   & 38.2588  &   1.090-11 \\
   63  &  2s2p$^5$3s		    &  $^4$P$^o_{1/2}$   & 	    & 	     	& 38.5119 &  38.4763   & 38.5271  &  38.5275   & 38.3978  &   1.457-11 \\
   64  &  2s2p$^5$($^3$P)3s	    &  $^2$P$^o_{3/2}$   & 38.657   & 	     	& 38.7074 &  38.6767   & 38.7624  &  38.7552   & 38.5795  &   2.206-12 \\
   65  &  2s2p$^5$($^3$P)3s	    &  $^2$P$^o_{1/2}$   & 38.858   & 	     	& 38.9127 &  38.8779   & 38.9622  &  38.9548   & 38.7779  &   1.786-12 \\
   66  &  2s2p$^5$3p		    &  $^4$S$  _{3/2}$   & 	    & 	     	& 39.6759 &  39.6458   & 39.6939  &  39.6906   & 39.5885  &   2.118-11 \\
   67  &  2s2p$^5$3p		    &  $^4$D$  _{7/2}$   & 	    & 	     	& 39.9070 &  39.8754   & 39.9253  &  39.9294   & 39.8137  &   2.538-11 \\
   68  &  2s2p$^5$3p		    &  $^4$D$  _{5/2}$   & 	    & 	     	& 39.9142 &  39.8845   & 39.9395  &  39.9432   & 39.8207  &   6.173-12 \\
   69  &  2s2p$^5$3p		    &  $^4$D$  _{3/2}$   & 	    & 	     	& 40.0059 &  39.9752   & 40.0297  &  40.0333   & 39.9103  &   5.149-12 \\
   70  &  2s2p$^5$($^3$P)3p	    &  $^2$D$  _{5/2}$   & 	    & 	     	& 40.0974 &  40.0672   & 40.1225  &  40.1258   & 39.9996  &   1.897-12 \\
   71  &  2s2p$^5$3p		    &  $^4$D$  _{1/2}$   & 	    & 	     	& 40.1030 &  40.0700   & 40.1348  &  40.1367   & 40.0040  &   1.119-11 \\
   72  &  2s2p$^5$($^3$P)3p	    &  $^2$P$  _{3/2}$   & 	    & 	     	& 40.2049 &  40.1709   & 40.2330  &  40.2379   & 40.1008  &   1.138-12 \\
   73  &  2s2p$^5$3p		    &  $^4$P$  _{5/2}$   & 	    & 	     	& 40.2454 &  40.2083   & 40.2694  &  40.2718   & 40.1391  &   2.380-12 \\
   74  &  2s2p$^5$3p		    &  $^4$P$  _{1/2}$   & 	    & 	     	& 40.2683 &  40.2337   & 40.2965  &  40.2950   & 40.1644  &   8.976-12 \\
   75  &  2s2p$^5$3p		    &  $^4$P$  _{3/2}$   & 	    & 	     	& 40.2727 &  40.2375   & 40.3026  &  40.3034   & 40.1676  &   2.442-12 \\
   76  &  2s2p$^5$($^3$P)3p	    &  $^2$P$  _{1/2}$   & 	    & 	     	& 40.3322 &  40.2980   & 40.3619  &  40.3654   & 40.2250  &   8.998-13 \\
   77  &  2s2p$^5$($^3$P)3p	    &  $^2$D$  _{3/2}$   & 	    & 	     	& 40.4258 &  40.3874   & 40.4508  &  40.4558   & 40.3139  &   1.344-12 \\
   78  &  2s2p$^5$($^1$P)3s	    &  $^2$P$^o_{3/2}$   & 	    & 	     	& 40.6851 &  40.6536   & 40.6805  &  40.6761   & 40.5309  &   2.445-12 \\
   79  &  2s2p$^5$($^1$P)3s	    &  $^2$P$^o_{1/2}$   & 	    & 	     	& 40.7039 &  40.6724   & 40.7017  &  40.6969   & 40.5482  &   3.133-12 \\
   80  &  2s2p$^5$($^3$P)3p	    &  $^2$S$  _{1/2}$   & 	    & 	     	& 40.7756 &  40.7424   & 40.8127  &  40.7916   & 40.6547  &   6.595-13 \\
\hline
\end{tabular}
\end{table}

\clearpage
\newpage
\setcounter{table}{0} 
\begin{table}
\caption{... continued.}
\small 
\centering
\begin{tabular}{rllrrrrrrrrr} \hline
Index  &     Configuration        & Level               & NIST  & HFR  &  GRASP1 & GRASP2        &   FAC1 & FAC2 & FAC3 & $\tau$  (GRASP2) \\
\hline
   81  &  2s2p$^5$3d		    &  $^4$P$^o_{1/2}$   & 	    & 	     	& 41.8397 &  41.8097   & 41.8679  &  41.8520   & 41.7237  &   2.127-11 \\
   82  &  2s2p$^5$3d		    &  $^4$P$^o_{3/2}$   & 	    & 	     	& 41.8813 &  41.8485   & 41.9081  &  41.8915   & 41.7630  &   1.966-11 \\
   83  &  2s2p$^5$3d		    &  $^4$F$^o_{9/2}$   & 	    & 	     	& 41.9310 &  41.8952   & 41.9715  &  41.9390   & 41.8241  &   3.837-11 \\
   84  &  2s2p$^5$3d		    &  $^4$P$^o_{5/2}$   & 	    & 	     	& 41.9563 &  41.9210   & 41.9821  &  41.9639   & 41.8356  &   2.265-11 \\
   85  &  2s2p$^5$3d		    &  $^4$F$^o_{7/2}$   & 	    & 	     	& 41.9942 &  41.9587   & 42.0314  &  41.9999   & 41.8824  &   3.677-11 \\
   86  &  2s2p$^5$3d		    &  $^4$F$^o_{5/2}$   & 	    & 	     	& 42.0775 &  42.0414   & 42.1135  &  42.0830   & 41.9626  &   3.450-11 \\
   87  &  2s2p$^5$3d		    &  $^4$F$^o_{3/2}$   & 	    & 	     	& 42.1521 &  42.1161   & 42.1905  &  42.1589   & 42.0387  &   1.909-11 \\
   88  &  2s2p$^5$3d		    &  $^4$D$^o_{7/2}$   & 	    & 	     	& 42.2121 &  42.1770   & 42.2404  &  42.2204   & 42.0850  &   2.897-11 \\
   89  &  2s2p$^5$($^1$P)3p	    &  $^2$D$  _{3/2}$   & 	    & 	     	& 42.2732 &  42.2418   & 42.2675  &  42.2594   & 42.1287  &   1.646-12 \\
   90  &  2s2p$^5$3d		    &  $^4$D$^o_{1/2}$   & 	    & 	     	& 42.3243 &  42.2883   & 42.3572  &  42.3311   & 42.2035  &   4.572-12 \\
   91  &  2s2p$^5$($^1$P)3p	    &  $^2$D$  _{5/2}$   & 	    & 	     	& 42.3436 &  42.3095   & 42.3363  &  42.3263   & 42.1962  &   1.957-12 \\
   92  &  2s2p$^5$3d		    &  $^4$D$^o_{5/2}$   & 	    & 	     	& 42.3491 &  42.3095   & 42.3746  &  42.3497   & 42.2201  &   2.657-11 \\
   93  &  2s2p$^5$3d		    &  $^4$D$^o_{3/2}$   & 	    & 	     	& 42.3633 &  42.3246   & 42.3937  &  42.3709   & 42.2401  &   7.444-12 \\
   94  &  2s2p$^5$($^3$P)3d	    &  $^2$F$^o_{7/2}$   & 	    & 	     	& 42.3700 &  42.3300   & 42.3888  &  42.3665   & 42.2308  &   2.915-11 \\
   95  &  2s2p$^5$($^3$P)3d	    &  $^2$D$^o_{5/2}$   & 	    & 	     	& 42.4019 &  42.3651   & 42.4224  &  42.4079   & 42.2585  &   2.755-11 \\
   96  &  2s2p$^5$($^1$P)3p	    &  $^2$P$  _{1/2}$   & 	    & 	     	& 42.4522 &  42.4212   & 42.4358  &  42.4416   & 42.3088  &   2.255-12 \\
   97  &  2s2p$^5$($^1$P)3p	    &  $^2$P$  _{3/2}$   & 	    & 	     	& 42.4859 &  42.4532   & 42.4688  &  42.4724   & 42.3407  &   2.618-12 \\
   98  &  2s2p$^5$($^3$P)3d	    &  $^2$D$^o_{3/2}$   & 	    & 	     	& 42.4946 &  42.4583   & 42.5160  &  42.5018   & 42.3507  &   2.525-12 \\
   99  &  2s2p$^5$($^3$P)3d	    &  $^2$F$^o_{5/2}$   & 	    & 	     	& 42.5757 &  42.5349   & 42.5907  &  42.5751   & 42.4246  &   2.676-11 \\
  100  &  2s2p$^5$($^1$P)3p	    &  $^2$S$  _{1/2}$   & 	    & 	     	& 42.6749 &  42.6456   & 42.7496  &  42.7062   & 42.5135  &   7.520-12 \\
  101  &  2s2p$^5$($^3$P)3d	    &  $^2$P$^o_{1/2}$   & 	    & 	     	& 42.7307 &  42.6969   & 42.7642  &  42.7443   & 42.5888  &   1.981-13 \\
  102  &  2s2p$^5$($^3$P)3d	    &  $^2$P$^o_{3/2}$   & 	    & 	     	& 42.9165 &  42.8747   & 42.9294  &  42.9221   & 42.7626  &   1.912-13 \\
\hline  
\end{tabular}
\begin{flushleft}
{\small
NIST:   {\tt http://www.nist.gov/pml/data/asd.cfm}  \\
HFR: Earlier results of Jup\'{e}n et al. \cite{cj} \\					      	       
GRASP1: Present results  with the {\sc grasp} code    for 501 level calculations {\em without} Breit and QED effects  \\	
GRASP2: Present results  with the {\sc grasp} code    for 501 level calculations {\em with} Breit and QED effects  \\			
FAC1: Present results  with the {\sc fac} code  for 113 level calculations  \\ 	
FAC2: Present results  with the {\sc fac} code  for 501 level calculations  \\ 
FAC3: Present results  with the {\sc fac} code  for 38~089 level calculations  \\ 
}
\end{flushleft}
\end{table}

\clearpage
\newpage

As with {\sc grasp}, with {\sc fac} too we have performed a series of calculations, but focus on only three, i.e. (i) FAC1,  which includes 113 levels as in GRASP1, (ii) FAC2, which includes 501 levels as in GRASP3, and finally (iii) FAC3, which includes in total 38~089 levels arising from all possible combinations of the (2*5) 3*2, 4*2, 5*2, 3*1 4*1, 3*1 5*1, and 4*1 5*1 configurations, plus those of FAC2. Although calculations have also been performed with even larger CI, these are not discussed here because the calculated energy levels show no appreciable differences, either in magnitude or orderings, i.e. the results have fully converged in FAC3.  For brevity, for {\sc fac} calculations we have used a short  notation here (and elsewhere in the text) for describing configurations. As an example,  3*2 means 3$\ell$3$\ell'$ resulting in 3s3p, 3s3d, 3p3d, 3s$^2$, 3p$^2$, and 3d$^2$.

Our calculated energies for the {\em lowest} 102 levels of Sc~XIII are listed in Table~1. These levels mostly belong to the 2s$^2$2p$^5$, 2s2p$^6$, 2s$^2$2p$^4$3$\ell$,  and 2s2p$^5$3$\ell$ configurations, and beyond these from others intermix, such as 2s$^2$2p$^4$4$\ell$.  However, energies for higher levels can be obtained from the author on request. Our energies calculated with {\sc grasp}, {\em without} and {\em with} the inclusion of Breit and QED (quantum electro-dynamic) effects, are listed in the table, along with all three calculations with the {\sc fac}, mentioned above. Also the experimental energies, compiled by the NIST (National Institute for Standards and Technology) team and available at the website {{\tt http://www.nist.gov/pml/data/asd.cfm}, are listed here along with the theoretical results of Jup\'{e}n et al. \cite{cj}, obtained from the Hartree-Fock Relativistic (HFR) code of Cowan -- see \cite{cow}. However, these theoretical results have been adjusted with {\em least square fitting} with the available measurements for a few levels, and that is the reason that there are no appreciable differences for the levels in common with the NIST. We also note that for two levels (40/41) the NIST and HFR energies are  non differentiable, but not in any of our calculations with both codes.

The contributions of the Breit and QED effects on the energy levels of Sc~XIII are not very significant, and are below 0.04~Ryd. However, these contributions have slightly lowered the energies and subsequently, discrepancies with those of NIST have increased, because comparatively there is a better match between the NIST and our energies obtained {\em without} them. Nevertheless, discrepancies between our results with {\sc grasp} (including the contributions from Breit and QED) and those of NIST are within 0.1~Ryd, and hence are highly satisfactory. Furthermore, there are no discrepancies in the level orderings between theory and measurements, and neither there are any ambiguities in level designations for this ion.

For most of the levels, there are no significant differences between the FAC1 and FAC2 energies, although the latter calculations include CI larger by more than a factor of four. However, for a few levels (such as 83 and 100) the differences are up to 0.04~Ryd, and energies in FAC2 are (mostly) lower. The same are the differences between the GRASP and FAC2 energies which include the same CI, but the latter ones are higher. Such small differences in energies between calculations with different codes are not uncommon and mainly arise due to the differences in algorithms, methodologies and formulations. Our FAC3 calculations include much larger CI and as a result the energies obtained are lower, by up to $\sim$0.2~Ryd, in comparison to those from FAC2. This has resulted in a better agreement with the GRASP energies. Although the FAC3 energies should be comparatively more accurate, differences with our GRASP or NIST are up to 0.25~Ryd -- see for example levels 45--49 and 56--58. Since the FAC3 energies are the lowest, we consider our results with {\sc grasp} to be comparatively more accurate, with agreement within 0.1~Ryd (0.3\%) with those of NIST, except for level 3 (2s2p$^6$~$^2$S$_{1/2}$) for which the discrepancy is 2\%, or 0.14~Ryd. For this level the energy calculated by J\"{o}nsson et al. \cite{jon} is closer to that of NIST, because not only they have included a significantly larger CI but their methodology is also different. Similarly,  combining CI with many-body perturbation theory (MBPT) approach, Gu \cite{gu} calculated the energy 6.945~Ryd, which is {\em lower} than that of NIST by only  0.014~Ryd, a tenth of the difference we have.

\begin{table}
\caption{Energies (in Ryd) and lifetimes ($\tau$, s) for the lowest 125 levels of  Sc XII. $a{\pm}b \equiv$ $a\times$10$^{{\pm}b}$.}
\small 
\centering
\begin{tabular}{rllrrrrr} \hline
Index  &     Configuration        & Level               & GRASP        &    FAC    &    $\tau$   (GRASP)     \\
 \hline		
    1   &   2s$^2$2p$^6$	  &  	  $^1$S$  _{0}$   &  0.0000  &  0.0000  &  ........	 \\ 
    2   &   2s$^2$2p$^5$3s	  &  	  $^3$P$^o_{2}$   & 29.3855  & 29.2066  &  4.462-05	 \\   
    3   &   2s$^2$2p$^5$3s	  &  	  $^3$P$^o_{1}$   & 29.4771  & 29.2951  &  4.228-12	 \\   
    4   &   2s$^2$2p$^5$3s	  &  	  $^3$P$^o_{0}$   & 29.7278  & 29.5422  &  1.530-03	 \\   
    5   &   2s$^2$2p$^5$3s	  &  	  $^1$P$^o_{1}$   & 29.8033  & 29.6144  &  2.977-12	 \\   
    6   &   2s$^2$2p$^5$3p	  &  	  $^3$S$  _{1}$   & 30.9051  & 30.7362  &  5.161-10	 \\ 
    7   &   2s$^2$2p$^5$3p	  &  	  $^3$D$  _{2}$   & 31.1455  & 30.9717  &  3.044-10	 \\ 
    8   &   2s$^2$2p$^5$3p	  &  	  $^3$D$  _{3}$   & 31.1574  & 30.9846  &  2.805-10	 \\ 
    9   &   2s$^2$2p$^5$3p	  &  	  $^3$D$  _{1}$   & 31.2401  & 31.0644  &  2.816-10	 \\ 
   10   &   2s$^2$2p$^5$3p	  &  	  $^3$P$  _{2}$   & 31.3076  & 31.1331  &  2.276-10	 \\ 
   11   &   2s$^2$2p$^5$3p	  &  	  $^1$P$  _{1}$   & 31.4772  & 31.2961  &  3.104-10	 \\ 
   12   &   2s$^2$2p$^5$3p	  &  	  $^3$P$  _{0}$   & 31.4838  & 31.3059  &  2.131-10	 \\ 
   13   &   2s$^2$2p$^5$3p	  &  	  $^3$P$  _{1}$   & 31.5815  & 31.4010  &  2.383-10	 \\ 
   14   &   2s$^2$2p$^5$3p	  &  	  $^1$D$  _{2}$   & 31.5802  & 31.3991  &  2.559-10	 \\ 
   15   &   2s$^2$2p$^5$3p	  &  	  $^1$S$  _{0}$   & 32.5413  & 32.3164  &  5.725-11	 \\ 
   16   &   2s$^2$2p$^5$3d	  &  	  $^3$P$^o_{0}$   & 33.3120  & 33.1018  &  1.287-10	 \\   
   17   &   2s$^2$2p$^5$3d	  &  	  $^3$P$^o_{1}$   & 33.3461  & 33.1362  &  3.549-11	 \\   
   18   &   2s$^2$2p$^5$3d	  &  	  $^3$P$^o_{2}$   & 33.4151  & 33.2043  &  1.303-10	 \\   
   19   &   2s$^2$2p$^5$3d	  &  	  $^3$F$^o_{4}$   & 33.4436  & 33.2375  &  1.278-10	 \\   
   20   &   2s$^2$2p$^5$3d	  &  	  $^3$F$^o_{3}$   & 33.4876  & 33.2737  &  1.190-10	 \\   
   21   &   2s$^2$2p$^5$3d	  &  	  $^3$F$^o_{2}$   & 33.5655  & 33.3496  &  1.142-10	 \\   
   22   &   2s$^2$2p$^5$3d	  &  	  $^3$D$^o_{3}$   & 33.6145  & 33.3953  &  1.166-10	 \\   
   23   &   2s$^2$2p$^5$3d	  &  	  $^3$D$^o_{1}$   & 33.7817  & 33.5589  &  1.357-12	 \\   
   24   &   2s$^2$2p$^5$3d	  &  	  $^1$D$^o_{2}$   & 33.8489  & 33.6321  &  1.131-10	 \\   
   25   &   2s$^2$2p$^5$3d	  &  	  $^3$D$^o_{2}$   & 33.8858  & 33.6592  &  1.151-10	 \\   
   26   &   2s$^2$2p$^5$3d	  &  	  $^1$F$^o_{3}$   & 33.8933  & 33.6665  &  1.194-10	 \\   
   27   &   2s$^2$2p$^5$3d	  &  	  $^1$P$^o_{1}$   & 34.2991  & 34.0503  &  1.331-13	 \\   
   28   &   2s2p$^6$3s  	  &  	  $^3$S$  _{1}$   & 36.5770  & 36.4646  &  1.004-11	 \\ 
   29   &   2s2p$^6$3s  	  &  	  $^1$S$  _{0}$   & 36.9177  & 36.7847  &  1.554-11	 \\ 
   30   &   2s2p$^6$3p  	  &  	  $^3$P$^o_{0}$   & 38.2806  & 38.1830  &  1.064-11	 \\   
   31   &   2s2p$^6$3p  	  &  	  $^3$P$^o_{1}$   & 38.2949  & 38.1970  &  6.973-12	 \\   
   32   &   2s2p$^6$3p  	  &  	  $^3$P$^o_{2}$   & 38.3440  & 38.2460  &  1.033-11	 \\   
   33   &   2s2p$^6$3p  	  &  	  $^1$P$^o_{1}$   & 38.4635  & 38.3613  &  8.246-13	 \\   
   34   &   2s$^2$2p$^5$4s	  &  	  $^3$P$^o_{2}$   & 39.4229  & 39.2385  &  9.358-12	 \\   
   35   &   2s$^2$2p$^5$4s	  &  	  $^1$P$^o_{1}$   & 39.4492  & 39.2665  &  4.976-12	 \\   
   36   &   2s$^2$2p$^5$4s	  &  	  $^3$P$^o_{0}$   & 39.7631  & 39.5717  &  9.131-12	 \\   
   37   &   2s$^2$2p$^5$4s	  &  	  $^3$P$^o_{1}$   & 39.7790  & 39.5889  &  5.632-12	 \\   
   38   &   2s$^2$2p$^5$4p	  &  	  $^3$S$  _{1}$   & 40.0442  & 39.8628  &  1.232-11	 \\ 
   39   &   2s$^2$2p$^5$4p	  &  	  $^3$D$  _{2}$   & 40.0825  & 39.9043  &  1.056-11	 \\ 
   40   &   2s$^2$2p$^5$4p	  &  	  $^3$D$  _{3}$   & 40.0825  & 39.9037  &  1.075-11	 \\ 
\hline
\end{tabular}
\end{table}

\clearpage
\newpage
\setcounter{table}{1} 
\begin{table}
\caption{... continued.}
\small 
\centering
\begin{tabular}{rllrrrrrrrrr} \hline
Index  &     Configuration        & Level               & GRASP        &    FAC    &    $\tau$   (GRASP)     \\
\hline
   41   &   2s$^2$2p$^5$4p	  &  	  $^1$P$  _{1}$   & 40.1279  & 39.9477  &  1.098-11	 \\ 
   42   &   2s$^2$2p$^5$4p	  &  	  $^3$P$  _{2}$   & 40.1497  & 39.9701  &  1.169-11	 \\ 
   43   &   2s$^2$2p$^5$4p	  &  	  $^3$P$  _{0}$   & 40.3069  & 40.1259  &  1.233-11	 \\ 
   44   &   2s$^2$2p$^5$4p	  &  	  $^3$D$  _{1}$   & 40.3850  & 40.2085  &  1.079-11	 \\ 
   45   &   2s$^2$2p$^5$4p	  &  	  $^1$D$  _{2}$   & 40.4298  & 40.2543  &  1.128-11	 \\ 
   46   &   2s$^2$2p$^5$4p	  &  	  $^3$P$  _{1}$   & 40.4623  & 40.2756  &  1.186-11	 \\ 
   47   &   2s2p$^6$3d  	  &  	  $^3$D$  _{3}$   & 40.5465  & 40.4081  &  2.170-11	 \\ 
   48   &   2s2p$^6$3d  	  &  	  $^3$D$  _{1}$   & 40.5616  & 40.4126  &  2.323-11	 \\ 
   49   &   2s2p$^6$3d  	  &  	  $^3$D$  _{2}$   & 40.5614  & 40.4136  &  2.152-11	 \\ 
   50   &   2s$^2$2p$^5$4p	  &  	  $^1$S$  _{0}$   & 40.7344  & 40.5575  &  1.427-11	 \\ 
   51   &   2s2p$^6$3d  	  &  	  $^1$D$  _{2}$   & 40.7676  & 40.6156  &  1.983-11	 \\ 
   52   &   2s$^2$2p$^5$4d	  &  	  $^3$P$^o_{0}$   & 40.8897  & 40.7058  &  7.852-12	 \\   
   53   &   2s$^2$2p$^5$4d	  &  	  $^3$P$^o_{1}$   & 40.9080  & 40.7234  &  6.962-12	 \\   
   54   &   2s$^2$2p$^5$4d	  &  	  $^3$F$^o_{4}$   & 40.9328  & 40.7469  &  7.748-12	 \\   
   55   &   2s$^2$2p$^5$4d	  &  	  $^3$P$^o_{2}$   & 40.9388  & 40.7529  &  7.898-12	 \\   
   56   &   2s$^2$2p$^5$4d	  &  	  $^3$F$^o_{3}$   & 40.9524  & 40.7648  &  7.871-12	 \\   
   57   &   2s$^2$2p$^5$4d	  &  	  $^1$D$^o_{2}$   & 40.9819  & 40.7928  &  8.035-12	 \\   
   58   &   2s$^2$2p$^5$4d	  &  	  $^3$D$^o_{3}$   & 40.9966  & 40.8076  &  8.035-12	 \\   
   59   &   2s$^2$2p$^5$4d	  &  	  $^3$D$^o_{1}$   & 41.1063  & 40.9116  &  8.331-13	 \\   
   60   &   2s$^2$2p$^5$4f	  &  	  $^3$G$  _{5}$   & 41.2824  & 41.0918  &  3.184-12	 \\ 
   61   &   2s$^2$2p$^5$4f	  &  	  $^1$G$  _{4}$   & 41.2829  & 41.0922  &  3.206-12	 \\ 
   62   &   2s$^2$2p$^5$4d	  &  	  $^3$F$^o_{2}$   & 41.2919  & 41.0986  &  7.908-12	 \\   
   63   &   2s$^2$2p$^5$4f	  &  	  $^1$F$  _{3}$   & 41.3078  & 41.1174  &  3.209-12	 \\ 
   64   &   2s$^2$2p$^5$4d	  &  	  $^3$D$^o_{2}$   & 41.3095  & 41.1062  &  3.215-12	 \\   
   65   &   2s$^2$2p$^5$4f	  &  	  $^3$F$  _{4}$   & 41.3006  & 41.1193  &  7.978-12	 \\ 
   66   &   2s$^2$2p$^5$4d	  &  	  $^1$F$^o_{3}$   & 41.3128  & 41.1177  &  7.893-12	 \\   
   67   &   2s$^2$2p$^5$4f	  &  	  $^3$F$  _{2}$   & 41.3262  & 41.1368  &  3.062-12	 \\ 
   68   &   2s$^2$2p$^5$4f	  &  	  $^3$F$  _{3}$   & 41.3284  & 41.1388  &  2.968-12	 \\ 
   69   &   2s$^2$2p$^5$4f	  &  	  $^3$D$  _{1}$   & 41.3386  & 41.1497  &  2.695-12	 \\ 
   70   &   2s$^2$2p$^5$4f	  &  	  $^3$D$  _{2}$   & 41.3388  & 41.1502  &  2.801-12	 \\ 
   71   &   2s$^2$2p$^5$4d	  &  	  $^1$P$^o_{1}$   & 41.4470  & 41.2417  &  3.844-13	 \\   
   72   &   2s$^2$2p$^5$4f	  &  	  $^3$G$  _{3}$   & 41.6359  & 41.4388  &  3.193-12	 \\ 
   73   &   2s$^2$2p$^5$4f	  &  	  $^3$G$  _{4}$   & 41.6383  & 41.4411  &  3.208-12	 \\ 
   74   &   2s$^2$2p$^5$4f	  &  	  $^1$D$  _{2}$   & 41.6614  & 41.4647  &  3.059-12	 \\ 
   75   &   2s$^2$2p$^5$4f	  &  	  $^3$D$  _{3}$   & 41.6638  & 41.4668  &  2.948-12	 \\ 
   76   &   2s$^2$2p$^5$5s	  &  	  $^3$P$^o_{2}$   & 43.6194  & 43.3963  &  1.163-11	 \\   
   77   &   2s$^2$2p$^5$5s	  &  	  $^1$P$^o_{1}$   & 43.6353  & 43.4126  &  6.884-12	 \\   
   78   &   2s$^2$2p$^5$5p	  &  	  $^3$S$  _{1}$   & 43.9292  & 43.7080  &  1.432-11	 \\ 
   79   &   2s$^2$2p$^5$5p	  &  	  $^3$D$  _{2}$   & 43.9556  & 43.7343  &  1.412-11	 \\ 
   80   &   2s$^2$2p$^5$5p	  &  	  $^3$D$  _{3}$   & 43.9580  & 43.7362  &  1.455-11	 \\ 
\hline
\end{tabular}
\end{table}

\clearpage
\newpage
\setcounter{table}{1} 
\begin{table}
\caption{... continued.}
\small 
\centering
\begin{tabular}{rllrrrrrrrrr} \hline
Index  &     Configuration        & Level               & GRASP        &    FAC    &    $\tau$   (GRASP)     \\
\hline
   81   &   2s$^2$2p$^5$5p	  &  	  $^1$P$  _{1}$   & 43.9727  & 43.7506  &  1.375-11	 \\ 
   82   &   2s$^2$2p$^5$5s	  &  	  $^3$P$^o_{0}$   & 43.9636  & 43.7332  &  1.161-11	 \\   
   83   &   2s$^2$2p$^5$5p	  &  	  $^3$P$  _{2}$   & 43.9820  & 43.7601  &  1.491-11	 \\ 
   84   &   2s$^2$2p$^5$5s	  &  	  $^3$P$^o_{1}$   & 43.9717  & 43.7416  &  7.711-12	 \\   
   85   &   2s$^2$2p$^5$5p	  &  	  $^3$P$  _{0}$   & 44.0864  & 43.8633  &  1.588-11	 \\ 
   86   &   2s$^2$2p$^5$5p	  &  	  $^3$D$  _{1}$   & 44.2937  & 44.0642  &  1.381-11	 \\ 
   87   &   2s$^2$2p$^5$5p	  &  	  $^3$P$  _{1}$   & 44.3076  & 44.0788  &  1.466-11	 \\ 
   88   &   2s$^2$2p$^5$5p	  &  	  $^1$D$  _{2}$   & 44.3137  & 44.0849  &  1.454-11	 \\ 
   89   &   2s$^2$2p$^5$5d	  &  	  $^3$P$^o_{0}$   & 44.3373  & 44.1172  &  9.963-12	 \\   
   90   &   2s$^2$2p$^5$5d	  &  	  $^3$P$^o_{1}$   & 44.3475  & 44.1267  &  9.039-12	 \\   
   91   &   2s$^2$2p$^5$5d	  &  	  $^3$F$^o_{4}$   & 44.3578  & 44.1361  &  1.015-11	 \\   
   92   &   2s$^2$2p$^5$5d	  &  	  $^3$P$^o_{2}$   & 44.3628  & 44.1408  &  1.013-11	 \\   
   93   &   2s$^2$2p$^5$5d	  &  	  $^3$F$^o_{3}$   & 44.3676  & 44.1450  &  1.022-11	 \\   
   94   &   2s$^2$2p$^5$5d	  &  	  $^1$D$^o_{2}$   & 44.3816  & 44.1577  &  1.038-11	 \\   
   95   &   2s$^2$2p$^5$5d	  &  	  $^3$D$^o_{3}$   & 44.3883  & 44.1642  &  1.034-11	 \\   
   96   &   2s$^2$2p$^5$5p	  &  	  $^1$S$  _{0}$   & 44.4089  & 44.1797  &  1.608-11	 \\ 
   97   &   2s$^2$2p$^5$5d	  &  	  $^3$D$^o_{1}$   & 44.4555  & 44.2249  &  9.363-13	 \\   
   98   &   2s$^2$2p$^5$5f	  &  	  $^3$D$  _{1}$   & 44.5338  & 44.3169  &  5.510-12	 \\ 
   99   &   2s$^2$2p$^5$5f	  &  	  $^3$G$  _{5}$   & 44.5342  & 44.3174  &  5.975-12	 \\ 
  100   &   2s$^2$2p$^5$5f	  &  	  $^1$G$  _{4}$   & 44.5348  & 44.3180  &  6.015-12	 \\ 
  101   &   2s$^2$2p$^5$5f	  &  	  $^3$D$  _{2}$   & 44.5363  & 44.3156  &  5.686-12	 \\ 
  102   &   2s$^2$2p$^5$5f	  &  	  $^3$F$  _{3}$   & 44.5440  & 44.3161  &  5.722-12	 \\ 
  103   &   2s$^2$2p$^5$5f	  &  	  $^3$F$  _{2}$   & 44.5458  & 44.3195  &  6.048-12	 \\ 
  104   &   2s$^2$2p$^5$5g	  &  	  $^3$F$^o_{2}$   & 44.5470  & 44.3291  &  1.129-11	 \\   
  105   &   2s$^2$2p$^5$5f	  &  	  $^1$F$  _{3}$   & 44.5470  & 44.3273  &  5.935-12	 \\ 
  106   &   2s$^2$2p$^5$5g	  &  	  $^3$F$^o_{3}$   & 44.5475  & 44.3231  &  1.131-11	 \\   
  107   &   2s$^2$2p$^5$5f	  &  	  $^3$F$  _{4}$   & 44.5479  & 44.3234  &  5.943-12	 \\ 
  108   &   2s$^2$2p$^5$5g	  &  	  $^1$H$^o_{5}$   & 44.5509  & 44.3193  &  1.129-11	 \\   
  109   &   2s$^2$2p$^5$5g	  &  	  $^3$H$^o_{6}$   & 44.5513  & 44.3197  &  1.129-11	 \\   
  110   &   2s$^2$2p$^5$5g	  &  	  $^3$G$^o_{3}$   & 44.5545  & 44.3301  &  1.133-11	 \\   
  111   &   2s$^2$2p$^5$5g	  &  	  $^3$G$^o_{4}$   & 44.5548  & 44.3260  &  1.132-11	 \\   
  112   &   2s$^2$2p$^5$5g	  &  	  $^1$G$^o_{4}$   & 44.5575  & 44.3313  &  1.132-11	 \\   
  113   &   2s$^2$2p$^5$5g	  &  	  $^3$G$^o_{5}$   & 44.5579  & 44.3264  &  1.133-11	 \\   
  114   &   2s$^2$2p$^5$5d	  &  	  $^3$F$^o_{2}$   & 44.7084  & 44.4787  &  1.026-11	 \\   
  115   &   2s$^2$2p$^5$5d	  &  	  $^3$D$^o_{2}$   & 44.7113  & 44.4816  &  1.019-11	 \\   
  116   &   2s$^2$2p$^5$5d	  &  	  $^1$F$^o_{3}$   & 44.7189  & 44.4885  &  1.021-11	 \\   
  117   &   2s$^2$2p$^5$5d	  &  	  $^1$P$^o_{1}$   & 44.7748  & 44.5376  &  8.700-13	 \\   
  118   &   2s$^2$2p$^5$5f	  &  	  $^3$G$  _{3}$   & 44.8831  & 44.6586  &  5.957-12	 \\ 
  119   &   2s$^2$2p$^5$5f	  &  	  $^3$G$  _{4}$   & 44.8846  & 44.6589  &  5.986-12	 \\ 
  120   &   2s$^2$2p$^5$5f	  &  	  $^3$D$  _{3}$   & 44.8870  & 44.6591  &  5.718-12	 \\ 
\hline
\end{tabular}
\end{table}

\clearpage
\newpage
\setcounter{table}{1} 
\begin{table}
\caption{... continued.}
\small 
\centering
\begin{tabular}{rllrrrrrrrrr} \hline
Index  &     Configuration        & Level               & GRASP        &    FAC    &    $\tau$   (GRASP)     \\
\hline
  122   &   2s$^2$2p$^5$5g	  &  	  $^1$F$^o_{3}$   & 44.8973  & 44.6628  &  1.133-11	 \\	 
  123   &   2s$^2$2p$^5$5g	  &  	  $^3$F$^o_{4}$   & 44.8977  & 44.6593  &  1.132-11	 \\	 
  124   &   2s$^2$2p$^5$5g	  &  	  $^3$H$^o_{4}$   & 44.8982  & 44.6604  &  1.130-11	 \\	 
  125   &   2s$^2$2p$^5$5g	  &  	  $^3$H$^o_{5}$   & 44.8987  & 44.6597  &  1.130-11	 \\ 
\hline  
\end{tabular}
\begin{flushleft}
{\small
GRASP: Present results  with the {\sc grasp} code  for 3948  level calculations  \\			
FAC: Present results  with the {\sc fac} code  for  93~437 level calculations  \\ 
}
\end{flushleft}
\end{table}

For most of the levels, there are no significant differences between the FAC1 and FAC2 energies, although the latter calculations include CI larger by more than a factor of four. However, for a few levels (such as 83 and 100) the differences are up to 0.04~Ryd, and energies in FAC2 are (mostly) lower. The same are the differences between the GRASP and FAC2 energies which include the same CI, but the latter ones are higher. Such small differences in energies between calculations with different codes are not uncommon and mainly arise due to the differences in algorithms, methodologies and formulations. Our FAC3 calculations include much larger CI and as a result the energies obtained are lower, by up to $\sim$0.2~Ryd, in comparison to those from FAC2. This has resulted in a better agreement with the GRASP energies. Although the FAC3 energies should be comparatively more accurate, differences with our GRASP or NIST are up to 0.25~Ryd -- see for example levels 45--49 and 56--58. Since the FAC3 energies are the lowest, we consider our results with {\sc grasp} to be comparatively more accurate, with agreement within 0.1~Ryd (0.3\%) with those of NIST, except for level 3 (2s2p$^6$~$^2$S$_{1/2}$) for which the discrepancy is 2\%, or 0.14~Ryd. For this level the energy calculated by J\"{o}nsson et al. \cite{jon} is closer to that of NIST, because not only they have included a significantly larger CI but their methodology is also different. Similarly,  combining CI with many-body perturbation theory (MBPT) approach, Gu \cite{gu} calculated the energy 6.945~Ryd, which is {\em lower} than that of NIST by only  0.014~Ryd, a tenth of the difference we have.

\subsection{Sc~XII}

As for Sc~XIII, for Sc~XII also we have performed a series of calculations with both {\sc grasp} and {\sc fac}. Our final calculations with {\sc grasp} include 3948 levels from 64 configurations, namely 2s$^2$2p$^6$, 2s$^2$2p$^5$$3\ell$, 2s2p$^6$$3\ell$,  2s$^2$2p$^5$4$\ell$,  2s2p$^6$$4\ell$,  2s$^2$2p$^5$5$\ell$, 2s2p$^6$$5\ell$, 2s$^2$2p$^5$6s/p/d, 2s$^2$2p$^5$7s/p/d, (2s$^2$2p$^4$) 3s3p, 3s3d, 3p3d, 3s$^2$, 3p$^2$, 3d$^2$, 3s4$\ell$, 3s5$\ell$, 3p4$\ell$,  3p5$\ell$, 3d4$\ell$, and 3d5$\ell$. Similarly, with {\sc fac} we have performed mainly three sets of calculations, which are: (i) FAC1, which includes 3948 levels of  the  same configurations as in GRASP,  (ii) FAC2, which includes 17~729 levels arising from all possible combinations of  2*8, (2*7) 3*1, 4*1, 5*1, 6*1, 7*1, (2*6) 3*2, 3*1 4*1, 3*1 5*1, 3*1 6*1, and 3*1 7*1,  and finally (iii) FAC3, which includes a total of 93~437 levels, the additional ones arising from  (2*6) 4*1 5*1, 4*1 6*1, 4*1 7*1, 5*1 6*1, 5*1 7*1, 6*1 7*1, and 2*5 3*3. These calculations are on the same lines as considered for some other Ne-like ions \cite{nelike1, nelike2}.

In Table~2, we list our {\em final} energies from both {\sc grasp} and {\sc fac} for the lowest 125 levels, because beyond these  from other configurations intermix, particularly from 2s$^2$2p$^5$6$\ell$.  However, energies for higher levels can be obtained from the author on request.  In general, energies obtained in FAC1 (not listed here but discussed below) are lower than of GRASP by $\sim$0.1~Ryd, and both calculations include the same CI. This observation is similar to that noted earlier for Sc~XIII. However, the FAC3 energies listed in Table~2 are lower than of GRASP by $\sim$0.2~Ryd, i.e. the effect of additional CI (by more than a factor of 20) is about 0.1~Ryd. In addition, for a few levels, such as 79--84, there are some (minor) differences in energy orderings, but overall there are no (major) discrepancies between calculations with two different codes. This result was expected and has been noted earlier for several ions, including some Ne-like \cite{nelike1, nelike2}, although some authors, such as \cite{sa}, have shown differences of up to $\sim$2~Ryd, but their calculations are incorrect as discussed in \cite{nelike1, nelike2} and further explained in \cite{nelike3, nelike4} -- see also \cite{rev2} for many other examples of discrepancies. Although a good agreement between the two calculations in our work confirms the accuracy of the calculated energies, we discuss these further below. 

As stated in Section~1, the only other energies available in the literature, but  only for the lowest 27 levels, are by Cogordan and Lunell \cite{cl} and J\"{o}nsson et al. \cite{jon}, who have also used (the different versions of) the {\sc grasp} code. Since experimental energies compiled by NIST are also available for a few levels of Sc~XII, in Table~3 we compare different sets of energies for the lowest 37 levels, which belong to the 2s$^2$2p$^6$, 2s$^2$2p$^5$3$\ell$,  2s2p$^6$3$\ell$, and 2s$^2$2p$^5$4s configurations. The FAC1 and FAC2 energies differ at most by 0.2~Ryd (see levels 29--33) which indicates a small effect of additional CI included in the latter. However, further inclusion of CI in FAC3 is not of any (great) advantage because differences with FAC2 are below 0.02~Ryd, i.e. the results have converged. However, energy differences between the FAC3 and NIST are the largest, and are up to 0.3~Ryd for several levels, and those from the former are invariably lower. Therefore as for Sc~XIII, energies calculated with {\sc fac} for Sc~XII too are comparatively less accurate. On the other hand, our calculations (and those of Cogordan and Lunell \cite{cl}) with {\sc grasp} compare well with those of NIST, because the differences are within $\sim$0.1~Ryd (0.3\%), with the measurements being (slightly) on the higher side. A notable exception is the level 15 (2s$^2$2p$^5$3p~$^1$S$_0$) for which  the energy calculated by Cogordan and Lunell  is (unusually) lower than our calculation by 0.14~Ryd. In all our calculations (with increasing CI) with the {\sc grasp} code the energy obtained for this level is invariably higher, and the contributions of Breit and QED effects are only 0.02~Ryd. Therefore, the reason for this (comparatively) large difference is neither in the inclusion of (much) larger CI in our calculations nor in the modified version of the code adopted, but is due to the fact that they have treated this level separately in a different manner. Anyway, their calculated energy for this level is as much lower than of NIST as ours is higher, and therefore the overall differences with measurements remain the same. Finally, the energies calculated by J\"{o}nsson et al. \cite{jon} are the most accurate because they have been able to produce results closer to those of NIST, for the same reasons as explained in Section~2.1 for Sc~XIII.

\begin{table}
\caption{Comparison of energies (in Ryd)  for the lowest 37 levels of  Sc XII.}
\small 
\centering
\begin{tabular}{rllrrrrrrrr} \hline
Index  &     Configuration        & Level                  & FAC1          & FAC2      & FAC3     & GRASP1   & GRASP2  & GRASP3 &  NIST       \\
 \hline
    1   &  2s$^2$2p$^6$ 	 &	 $^1$S$  _{0}$     &	0.0000     &  0.0000  &  0.0000 &  0.0000  &   0.0000  &    0.0000 &  0.0000   \\
    2   &  2s$^2$2p$^5$3s	 &	 $^3$P$^o_{2}$     &   29.2619     & 29.2128  & 29.2066 & 29.4042  &  29.4787  &   29.3855 & 29.4811   \\
    3   &  2s$^2$2p$^5$3s	 &	 $^3$P$^o_{1}$     &   29.3592     & 29.3010  & 29.2951 & 29.5004  &  29.5695  &   29.4771 & 29.5720   \\
    4   &  2s$^2$2p$^5$3s	 &	 $^3$P$^o_{0}$     &   29.5982     & 29.5483  & 29.5422 & 29.7468  &  29.8215  &   29.7278 & 29.8233   \\
    5   &  2s$^2$2p$^5$3s	 &	 $^1$P$^o_{1}$     &   29.6831     & 29.6201  & 29.6144 & 29.8293  &  29.8956  &   29.8033 & 29.8970   \\
    6   &  2s$^2$2p$^5$3p	 &	 $^3$S$  _{1}$     &   30.8156     & 30.7434  & 30.7362 & 30.9214  &  30.9967  &   30.9051 & 30.9960   \\
    7   &  2s$^2$2p$^5$3p	 &	 $^3$D$  _{2}$     &   31.0546     & 30.9774  & 30.9717 & 31.1597  &  31.2317  &   31.1455 & 31.2312   \\
    8   &  2s$^2$2p$^5$3p	 &	 $^3$D$  _{3}$     &   31.0678     & 30.9905  & 30.9846 & 31.1703  &  31.2434  &   31.1574 & 31.2431   \\
    9   &  2s$^2$2p$^5$3p	 &	 $^3$D$  _{1}$     &   31.1490     & 31.0702  & 31.0644 & 31.2538  &  31.3235  &   31.2401 & 31.3234   \\
   10   &  2s$^2$2p$^5$3p	 &	 $^3$P$  _{2}$     &   31.2157     & 31.1387  & 31.1331 & 31.3230  &  31.3942  &   31.3076 & 31.3940   \\
   11   &  2s$^2$2p$^5$3p	 &	 $^1$P$  _{1}$     &   31.3817     & 31.3018  & 31.2961 & 31.4906  &  31.5606  &   31.4772 & 31.5598   \\
   12   &  2s$^2$2p$^5$3p	 &	 $^3$P$  _{0}$     &   31.3885     & 31.3117  & 31.3059 & 31.4988  &  31.5678  &   31.4838 & 31.5681   \\
   13   &  2s$^2$2p$^5$3p	 &	 $^3$P$  _{1}$     &   31.4852     & 31.4067  & 31.4010 & 31.5960  &  31.6668  &   31.5815 & 31.6662   \\
   14   &  2s$^2$2p$^5$3p	 &	 $^1$D$  _{2}$     &   31.4831     & 31.4046  & 31.3991 & 31.5952  &  31.6669  &   31.5802 & 31.6662   \\
   15   &  2s$^2$2p$^5$3p	 &	 $^1$S$  _{0}$     &   32.4021     & 32.3332  & 32.3164 & 32.4017  &  32.4890  &   32.5413 & 32.4845   \\
   16   &  2s$^2$2p$^5$3d	 &	 $^3$P$^o_{0}$     &   33.1852     & 33.1085  & 33.1018 & 33.3191  &  33.3960  &   33.3120 &	       \\
   17   &  2s$^2$2p$^5$3d	 &	 $^3$P$^o_{1}$     &   33.2200     & 33.1429  & 33.1362 & 33.3534  &  33.4287  &   33.3461 & 33.4302   \\
   18   &  2s$^2$2p$^5$3d	 &	 $^3$P$^o_{2}$     &   33.2889     & 33.2110  & 33.2043 & 33.4229  &  33.4966  &   33.4151 & 33.4980   \\
   19   &  2s$^2$2p$^5$3d	 &	 $^3$F$^o_{4}$     &   33.3187     & 33.2445  & 33.2375 & 33.4556  &  33.5222  &   33.4436 & 33.5237   \\
   20   &  2s$^2$2p$^5$3d	 &	 $^3$F$^o_{3}$     &   33.3577     & 33.2807  & 33.2737 & 33.5008  &  33.5636  &   33.4876 & 33.5650   \\
   21   &  2s$^2$2p$^5$3d	 &	 $^3$F$^o_{2}$     &   33.4356     & 33.3565  & 33.3496 & 33.5772  &  33.6396  &   33.5655 & 33.6413   \\
   22   &  2s$^2$2p$^5$3d	 &	 $^3$D$^o_{3}$     &   33.4844     & 33.4020  & 33.3953 & 33.6274  &  33.6874  &   33.6145 & 33.6887   \\
   23   &  2s$^2$2p$^5$3d	 &	 $^3$D$^o_{1}$     &   33.6480     & 33.5660  & 33.5589 & 33.7940  &  33.8517  &   33.7817 & 33.8510   \\
   24   &  2s$^2$2p$^5$3d	 &	 $^1$D$^o_{2}$     &   33.7188     & 33.6389  & 33.6321 & 33.8605  &  33.9227  &   33.8489 & 33.9238   \\
   25   &  2s$^2$2p$^5$3d	 &	 $^3$D$^o_{2}$     &   33.7487     & 33.6658  & 33.6592 & 33.8458  &  33.9598  &   33.8858 & 33.9608   \\
   26   &  2s$^2$2p$^5$3d	 &	 $^1$F$^o_{3}$     &   33.7549     & 33.6732  & 33.6665 & 33.9062  &  33.9674  &   33.8933 & 33.9682   \\
   27   &  2s$^2$2p$^5$3d	 &	 $^1$P$^o_{1}$     &   34.1446     & 34.0606  & 34.0503 & 34.3009  &  34.3355  &   34.2991 & 34.3300   \\
   28   &  2s2p$^6$3s		 &	 $^3$S$  _{1}$     &   36.6268     & 36.4760  & 36.4646 & 	   &  	       &   36.5770 &	       \\
   29   &  2s2p$^6$3s		 &	 $^1$S$  _{0}$     &   36.9957     & 36.7993  & 36.7847 & 	   &  	       &   36.9177 &	       \\
   30   &  2s2p$^6$3p		 &	 $^3$P$^o_{0}$     &   38.3366     & 38.1944  & 38.1830 & 	   &  	       &   38.2806 &	       \\
   31   &  2s2p$^6$3p		 &	 $^3$P$^o_{1}$     &   38.3507     & 38.2084  & 38.1970 & 	   &  	       &   38.2949 & 38.2550   \\
   32   &  2s2p$^6$3p		 &	 $^3$P$^o_{2}$     &   38.3981     & 38.2575  & 38.2460 & 	   &  	       &   38.3440 &	       \\
   33   &  2s2p$^6$3p		 &	 $^1$P$^o_{1}$     &   38.5280     & 38.3721  & 38.3613 & 	   &  	       &   38.4635 & 38.4100   \\
   34   &  2s$^2$2p$^5$4s	 &	 $^3$P$^o_{2}$     &   39.3200     & 39.2610  & 39.2385 & 	   &  	       &   39.4229 & 39.5248   \\
   35   &  2s$^2$2p$^5$4s	 &	 $^1$P$^o_{1}$     &   39.3463     & 39.2892  & 39.2665 & 	   &  	       &   39.4492 & 39.5430   \\
   36   &  2s$^2$2p$^5$4s	 &	 $^3$P$^o_{0}$     &   39.6538     & 39.5950  & 39.5717 & 	   &  	       &   39.7631 &	       \\
   37   &  2s$^2$2p$^5$4s	 &	 $^3$P$^o_{1}$     &   39.6702     & 39.6123  & 39.5889 & 	   &  	       &   39.7790 & 39.9030   \\
\hline				  				        			
\end{tabular}										             				    
\begin {flushleft}									            			    
\begin{tabbing} 									      	
aaaaaaaaaaaaaaaaaaaaaaaaaaaaaaaaaaaa\= \kill						      	       
FAC1: Present results  with the {\sc fac} code for 3948   level calculations \\
FAC2: Present results  with the {\sc fac} code for 17~729 level calculations  \\		
FAC3: Present results  with the {\sc fac} code for  93~437 level calculations  \\ 
GRASP1: Earlier results of Cogordan and Lunell  \cite{cl} with the {\sc grasp} code \\
GRASP2: Earlier results of J\"{o}nsson et al. \cite{jon} with the {\sc grasp} code \\	
GRASP3: Present results  with the {\sc grasp} code for 3948   level calculations \\
NIST:   {\tt http://www.nist.gov/pml/data/asd.cfm}  \\ 	 			      		      	
\end{tabbing}										      	
\end {flushleft}				
\end{table}

\subsection{Y~XXX}

For Y~XXX we have performed similar calculations with {\sc grasp} and {\sc fac} as for Sc~XII, described in Section~2.2. In Table~4 we list our {\em final} results with both these codes for the lowest 139 levels, as beyond these is an intermix from other configurations, such as 2s$^2$2p$^5$6$\ell$.  However, energies for higher levels can be obtained from the author on request.  As stated in Section~1, similar results for some levels are available in the literature by Zhang and Sampson \cite{zs2} and  Hagelstein and Jung \cite{hag}, who have used the Dirac-Fock-Slater (DFS) and YODA codes, respectively.  The GRASP results are listed with inclusion (GRASP1) and exclusion (GRASP2)  of Breit and QED effects, because Y~XXX is comparatively a heavy ion, and their net effect is to lower the energies by a maximum of 0.4~Ryd. Their effect is comparatively more noticeable on higher levels than the lower ones, and the maximum is on the ground level, i.e. 5.60~Ryd (3.25 + 2.35). The DFS energies, available for 89 levels, are closer to the GRASP2 results, which indicates the neglect of higher order relativistic effects from the calculations. The other results from YODA, although for only the lowest 37 levels, are closer to GRASP1 and agree within 0.1~Ryd, which is highly satisfactory. However, as for other two ions, the corresponding results with {\sc fac} are the lowest among those listed in Table~4, although differ by a maximum of  0.2~Ryd with GRASP1. Before drawing our conclusion we make a few other comparisons below.

In Table~5 we compare our GRASP results with those of Cogordan and Lunell \cite{cl} and Quinet et al. \cite{pq}, who have used the same code but of different versions. Also included in the table are our results from three calculations with {\sc fac}, with increasing CI. All three sets of energies agree within 0.2~Ryd (see level 29) and hence the differences are insignificant, i.e. $<$0.1\%. Similarly, all calculations with {\sc grasp} agree within 0.1~Ryd and hence provide confidence in our results listed in Table~4. Since the FAC energies are the lowest, irrespective of the level of CI, these results are assessed to be comparatively less accurate, and therefore our results obtained with the {\sc grasp}  should be considered to be more accurate, and perhaps the best available to date for a larger number of levels. Further accuracy of our energy levels can be confirmed by the measurements which unfortunately are not yet available for the levels of Y~XXX.

\section{Radiative rates}

Our results for the $A$-values calculated with the {\sc grasp} code are listed in Tables~6--8 for the transitions in Sc~XIII, Sc~XII and Y~XXX, respectively.  For brevity only resonance transitions, i.e. from the ground level, are listed here, but complete results for {\em all} transitions in ASCII format are available  online as a supplementary material, and the indices for the lower ($i$) and upper ($j$) levels correspond to those listed in Tables~1, 2 and 4, for the respective ions. Furthermore, for the E1 transitions we list absorption oscillator strength ($f_{ij}$) and  line strength $S$ (in atomic unit, 1 a.u. = 6.460$\times$10$^{-36}$ cm$^2$ esu$^2$) apart from the $A$-values, but only the latter parameter for other types of transitions, i.e. E2, M1 and M2. However, desired results for $f$- or $S$-values for these transitions can be easily obtained from the standard equations which have been listed in some of our earlier papers, but are also given below for a ready reference, i.e.

\begin{flushleft}
for the electric dipole (E1) transitions 
\end{flushleft} 
\begin{equation}
A_{ji} = \frac{2.0261\times{10^{18}}}{{{\omega}_j}\lambda^3_{ji}} S^{E1} \hspace*{0.5 cm} {\rm and} \hspace*{0.5 cm} 
f_{ij} = \frac{303.75}{\lambda_{ji}\omega_i} S^{E1}, \\
\end{equation}
\begin{flushleft}
for the magnetic dipole (M1) transitions  
\end{flushleft}
\begin{equation}
A_{ji} = \frac{2.6974\times{10^{13}}}{{{\omega}_j}\lambda^3_{ji}} S^{M1} \hspace*{0.5 cm} {\rm and} \hspace*{0.5 cm}
f_{ij} = \frac{4.044\times{10^{-3}}}{\lambda_{ji}\omega_i} S^{M1}, \\
\end{equation}
\begin{flushleft}
for the electric quadrupole (E2) transitions
\end{flushleft}
\begin{equation}
A_{ji} = \frac{1.1199\times{10^{18}}}{{{\omega}_j}\lambda^5_{ji}} S^{E2} \hspace*{0.5 cm} {\rm and} \hspace*{0.5 cm}
f_{ij} = \frac{167.89}{\lambda^3_{ji}\omega_i} S^{E2}, 
\end{equation}
\begin{flushleft}
and for the magnetic quadrupole (M2) transitions 
\end{flushleft}
\begin{equation}
A_{ji} = \frac{1.4910\times{10^{13}}}{{{\omega}_j}\lambda^5_{ji}} S^{M2} \hspace*{0.5 cm} {\rm and} \hspace*{0.5 cm}
f_{ij} = \frac{2.236\times{10^{-3}}}{\lambda^3_{ji}\omega_i} S^{M2}. \\
\end{equation}

\begin{flushleft} 
We also note here that $f$- and $A$-values are related as 
\end{flushleft}
\begin{equation}
f_{ij} = \frac{mc}{8{\pi}^2{e^2}}{\lambda^2_{ji}} \frac{{\omega}_j}{{\omega}_i}A_{ji}
 = 1.49 \times 10^{-16} \lambda^2_{ji} \frac{{\omega}_j}{{\omega}_i} A_{ji}
\end{equation}
where $m$ and $e$ are the electron mass and charge, respectively, $c$ is the velocity of light, 
$\lambda_{ji}$ is the transition wavelength in \AA, and $\omega_i$ and $\omega_j$ are the statistical weights of the lower $i$ and upper $j$ levels, respectively. This relationship is the same irrespective of the type of a transition, and $\lambda_{ji}$ are also listed in Tables~6--8 for all possible transitions.

\begin{table}
\caption{Energies (in Ryd) and lifetimes ($\tau$, s) for the lowest 139 levels of  Y XXX. $a{\pm}b \equiv$ $a\times$10$^{{\pm}b}$.}
\small 
\centering
\begin{tabular}{rllrrrrrrrr} \hline
Index  &     Configuration        & Level                  & GRASP1 & GRASP2        & FAC  &  DFS    &  YODA     &       $\tau$    (GRASP1)    \\
 \hline	
    1   &   2s$^2$2p$^6$	  &  	  $^1$S$  _{0}$    &	 0.0000    &   0.0000  &   0.0000   &	 0.0000 &   0.0000  &	    ........   \\
    2   &   2s$^2$2p$^5$3s	  &  	  $^3$P$^o_{2}$    &   146.6987    & 146.8657  & 146.5497   &  146.8576 & 146.8047  &	    8.827-08   \\
    3   &   2s$^2$2p$^5$3s	  &  	  $^1$P$^o_{1}$    &   146.9606    & 147.1241  & 146.8078   &  147.1295 & 147.0678  &	    1.283-13   \\
    4   &   2s$^2$2p$^5$3p	  &  	  $^3$S$  _{1}$    &   150.9658    & 151.1208  & 150.8223   &  151.0984 & 151.0374  &	    1.067-10   \\
    5   &   2s$^2$2p$^5$3p	  &  	  $^3$D$  _{2}$    &   151.1804    & 151.3472  & 151.0345   &  151.3336 & 151.2476  &	    4.442-11   \\
    6   &   2s$^2$2p$^5$3p	  &  	  $^3$D$  _{3}$    &   152.4685    & 152.6663  & 152.3230   &  152.6493 & 152.5405  &	    4.083-11   \\
    7   &   2s$^2$2p$^5$3p	  &  	  $^1$P$  _{1}$    &   152.4932    & 152.6816  & 152.3437   &  152.6640 & 152.5647  &	    4.749-11   \\
    8   &   2s$^2$2p$^5$3s	  &  	  $^3$P$^o_{0}$    &   152.4791    & 152.7338  & 152.3099   &  152.7375 & 152.6132  &	    1.805-07   \\
    9   &   2s$^2$2p$^5$3s	  &  	  $^3$P$^o_{1}$    &   152.6077    & 152.8713  & 152.4363   &  152.8771 & 152.7433  &	    1.889-13   \\
   10   &   2s$^2$2p$^5$3p	  &  	  $^3$P$  _{2}$    &   152.8584    & 153.0399  & 152.7116   &  153.0314 & 152.9293  &	    2.449-11   \\
   11   &   2s$^2$2p$^5$3p	  &  	  $^3$P$  _{0}$    &   154.2837    & 154.4569  & 154.1154   &  154.4647 & 154.3669  &	    1.989-11   \\
   12   &   2s$^2$2p$^5$3p	  &  	  $^3$D$  _{1}$    &   156.8517    & 157.1167  & 156.6846   &  157.1106 & 156.9475  &	    1.010-10   \\
   13   &   2s$^2$2p$^5$3d	  &  	  $^3$P$^o_{0}$    &   157.6799    & 157.8546  & 157.5000   &  157.8309 & 157.7471  &	    3.221-11   \\
   14   &   2s$^2$2p$^5$3d	  &  	  $^3$P$^o_{1}$    &   157.8890    & 158.0863  & 157.7075   &  158.0661 & 157.9573  &	    3.807-12   \\
   15   &   2s$^2$2p$^5$3d	  &  	  $^3$F$^o_{3}$    &   158.1738    & 158.3932  & 157.9895   &  158.3822 & 158.2447  &	    2.864-11   \\
   16   &   2s$^2$2p$^5$3d	  &  	  $^3$D$^o_{2}$    &   158.2712    & 158.4762  & 158.0866   &  158.4557 & 158.3395  &	    3.116-11   \\
   17   &   2s$^2$2p$^5$3d	  &  	  $^3$F$^o_{4}$    &   158.2876    & 158.5230  & 158.1143   &  158.5071 & 158.3461  &	    4.574-11   \\
   18   &   2s$^2$2p$^5$3p	  &  	  $^3$P$  _{1}$    &   158.3146    & 158.5880  & 158.1492   &  158.5806 & 158.4160  &	    3.794-11   \\
   19   &   2s$^2$2p$^5$3p	  &  	  $^1$D$  _{2}$    &   158.4251    & 158.7151  & 158.2578   &  158.7129 & 158.5262  &	    2.562-11   \\
   20   &   2s$^2$2p$^5$3d	  &  	  $^1$D$^o_{2}$    &   158.5283    & 158.7396  & 158.3391   &  158.7202 & 158.6004  &	    3.485-11   \\
   21   &   2s$^2$2p$^5$3p	  &  	  $^1$S$  _{0}$    &   158.6252    & 158.8512  & 158.4291   &  158.8893 & 158.7342  &	    1.943-11   \\
   22   &   2s$^2$2p$^5$3d	  &  	  $^3$D$^o_{3}$    &   158.8049    & 159.0240  & 158.6145   &  159.0142 & 158.8812  &	    4.014-11   \\
   23   &   2s$^2$2p$^5$3d	  &  	  $^3$D$^o_{1}$    &   159.7700    & 159.9925  & 159.5589   &  159.9991 & 159.8683  &	    8.547-15   \\
   24   &   2s$^2$2p$^5$3d	  &  	  $^3$F$^o_{2}$    &   163.9104    & 164.2167  & 163.7151   &  164.2106 & 164.0077  &	    2.803-11   \\
   25   &   2s$^2$2p$^5$3d	  &  	  $^3$P$^o_{2}$    &   164.2275    & 164.5414  & 164.0182   &  164.5340 & 164.3245  &	    4.173-11   \\
   26   &   2s$^2$2p$^5$3d	  &  	  $^1$F$^o_{3}$    &   164.3548    & 164.6758  & 164.1453   &  164.6736 & 164.4495  &	    4.267-11   \\
   27   &   2s$^2$2p$^5$3d	  &  	  $^1$P$^o_{1}$    &   164.8886    & 165.2049  & 164.6642   &  165.2175 & 164.9992  &	    7.820-15   \\
   28   &   2s2p$^6$3s  	  &  	  $^3$S$  _{1}$    &   167.5933    & 167.8473  & 167.5051   &  167.8635 & 167.6790  &	    2.216-12   \\
   29   &   2s2p$^6$3s  	  &  	  $^1$S$  _{0}$    &   168.4649    & 168.7070  & 168.3525   &  168.7381 & 168.5500  &	    3.028-12   \\
   30   &   2s2p$^6$3p  	  &  	  $^3$P$^o_{0}$    &   171.9546    & 172.1943  & 171.8826   &  172.2072 & 172.0073  &	    2.413-12   \\
   31   &   2s2p$^6$3p  	  &  	  $^3$P$^o_{1}$    &   172.0393    & 172.2913  & 171.9655   &  172.3028 & 172.0918  &	    1.180-13   \\
   32   &   2s2p$^6$3p  	  &  	  $^3$P$^o_{2}$    &   173.3963    & 173.6725  & 173.3240   &  173.6846 & 173.4560  &	    2.304-12   \\
   33   &   2s2p$^6$3p  	  &  	  $^1$P$^o_{1}$    &   173.6114    & 173.8890  & 173.5349   &  173.9051 & 173.6669  &	    3.951-14   \\
   34   &   2s2p$^6$3d  	  &  	  $^3$D$  _{1}$    &   178.8170    & 179.0985  & 178.7101   &  179.1014 & 178.8809  &	    2.685-12   \\
   35   &   2s2p$^6$3d  	  &  	  $^3$D$  _{2}$    &   178.8986    & 179.1968  & 178.7914   &  179.1970 & 178.9632  &	    2.541-12   \\
   36   &   2s2p$^6$3d  	  &  	  $^3$D$  _{3}$    &   179.1053    & 179.4192  & 178.9983   &  179.4248 & 179.1624  &	    2.585-12   \\
   37   &   2s2p$^6$3d  	  &  	  $^1$D$  _{2}$    &   179.7476    & 180.0497  & 179.6257   &  180.0789 & 179.8210  &	    1.706-12   \\
   38   &   2s$^2$2p$^5$4s	  &  	  $^3$P$^o_{2}$    &   199.3165    & 199.5290  & 199.1716   &  199.5414 &	    &	    3.442-13   \\
   39   &   2s$^2$2p$^5$4s	  &  	  $^1$P$^o_{1}$    &   199.4034    & 199.6142  & 199.2606   &  199.6296 &	    &	    1.803-13   \\
   40   &   2s$^2$2p$^5$4p	  &  	  $^3$S$  _{1}$    &   201.0886    & 201.2950  & 200.9476   &  201.3053 &	    &	    3.072-13   \\
\hline
\end{tabular}
\end{table}

\clearpage
\newpage
\setcounter{table}{3} 
\begin{table}
\caption{... continued.}
\small 
\centering
\begin{tabular}{rllrrrrrrrrr} \hline
Index  &     Configuration        & Level                  & GRASP1 & GRASP2        & FAC  &  DFS    &  YODA     &       $\tau$    (GRASP1)    \\
\hline
   41   &   2s$^2$2p$^5$4p	  &  	  $^3$D$  _{2}$    &   201.1477    & 201.3587  & 201.0071   &  201.3641 &	    &	    3.033-13   \\
   42   &   2s$^2$2p$^5$4p	  &  	  $^3$D$  _{3}$    &   201.6754    & 201.8994  & 201.5336   &  201.9080 &	    &	    3.378-13   \\
   43   &   2s$^2$2p$^5$4p	  &  	  $^1$P$  _{1}$    &   201.6926    & 201.9130  & 201.5496   &  201.9227 &	    &	    3.317-13   \\
   44   &   2s$^2$2p$^5$4p	  &  	  $^3$P$  _{2}$    &   201.8149    & 202.0323  & 201.6736   &  202.0403 &	    &	    3.445-13   \\
   45   &   2s$^2$2p$^5$4p	  &  	  $^1$S$  _{0}$    &   202.4018    & 202.6089  & 202.2643   &  202.6210 &	    &	    3.651-13   \\
   46   &   2s$^2$2p$^5$4d	  &  	  $^3$P$^o_{0}$    &   203.6528    & 203.8698  & 203.5107   &  203.8851 &	    &	    1.740-13   \\
   47   &   2s$^2$2p$^5$4d	  &  	  $^3$P$^o_{1}$    &   203.7330    & 203.9577  & 203.5899   &  203.9733 &	    &	    1.709-13   \\
   48   &   2s$^2$2p$^5$4d	  &  	  $^3$F$^o_{3}$    &   203.8229    & 204.0551  & 203.6770   &  204.0615 &	    &	    1.763-13   \\
   49   &   2s$^2$2p$^5$4d	  &  	  $^3$D$^o_{2}$    &   203.8692    & 204.0950  & 203.7236   &  204.1056 &	    &	    1.763-13   \\
   50   &   2s$^2$2p$^5$4d	  &  	  $^3$F$^o_{4}$    &   203.8863    & 204.1251  & 203.7419   &  204.1350 &	    &	    1.740-13   \\
   51   &   2s$^2$2p$^5$4d	  &  	  $^1$D$^o_{2}$    &   203.9686    & 204.1992  & 203.8208   &  204.2085 &	    &	    1.754-13   \\
   52   &   2s$^2$2p$^5$4d	  &  	  $^3$D$^o_{3}$    &   204.0699    & 204.3017  & 203.9220   &  204.3114 &	    &	    1.762-13   \\
   53   &   2s$^2$2p$^5$4d	  &  	  $^1$P$^o_{1}$    &   204.4392    & 204.6729  & 204.2800   &  204.6789 &	    &	    1.619-14   \\
   54   &   2s$^2$2p$^5$4f	  &  	  $^3$D$  _{1}$    &   205.0260    & 205.2616  & 204.8713   &  205.2816 &	    &	    8.126-14   \\
   55   &   2s$^2$2p$^5$4f	  &  	  $^3$G$  _{4}$    &   205.0516    & 205.2961  & 204.8970   &  205.3184 &	    &	    8.452-14   \\
   56   &   2s$^2$2p$^5$4f	  &  	  $^3$D$  _{2}$    &   205.0755    & 205.3135  & 204.9208   &  205.3331 &	    &	    8.218-14   \\
   57   &   2s$^2$2p$^5$4f	  &  	  $^3$G$  _{5}$    &   205.0973    & 205.3419  & 204.9425   &  205.3625 &	    &	    8.462-14   \\
   58   &   2s$^2$2p$^5$4f	  &  	  $^3$F$  _{3}$    &   205.1329    & 205.3729  & 204.9787   &  205.3919 &	    &	    8.392-14   \\
   59   &   2s$^2$2p$^5$4f	  &  	  $^1$D$  _{2}$    &   205.1513    & 205.3899  & 204.9965   &  205.4139 &	    &	    8.569-14   \\
   60   &   2s$^2$2p$^5$4f	  &  	  $^1$F$  _{3}$    &   205.1659    & 205.4060  & 205.0108   &  205.4246 &	    &	    8.382-14   \\
   61   &   2s$^2$2p$^5$4s	  &  	  $^3$P$^o_{0}$    &   205.1153    & 205.4217  & 204.9493   &  205.4433 &   &	    3.427-13   \\
   62   &   2s$^2$2p$^5$4f	  &  	  $^3$F$  _{4}$    &   205.1953    & 205.4368  & 205.0410   &  205.4580 &   &	    8.490-14   \\
   63   &   2s$^2$2p$^5$4s	  &  	  $^3$P$^o_{1}$    &   205.1705    & 205.4783  & 205.0058   &  205.5021 &   &	    3.381-13   \\
   64   &   2s$^2$2p$^5$4p	  &  	  $^3$D$  _{1}$    &   206.9038    & 207.2126  & 206.7408   &  207.2293 &   &	    2.999-13   \\
   65   &   2s$^2$2p$^5$4p	  &  	  $^3$P$  _{0}$    &   207.3845    & 207.6802  & 207.2255   &  207.7071 &   &	    3.224-13   \\
   66   &   2s$^2$2p$^5$4p	  &  	  $^3$P$  _{1}$    &   207.4992    & 207.8124  & 207.3363   &  207.8320 &   &	    3.411-13   \\
   67   &   2s$^2$2p$^5$4p	  &  	  $^1$D$  _{2}$    &   207.5333    & 207.8527  & 207.3708   &  207.8688 &   &	    3.384-13   \\
   68   &   2s$^2$2p$^5$4d	  &  	  $^3$F$^o_{2}$    &   209.5980    & 209.9242  & 209.4319   &  209.9488 &   &	    1.763-13   \\
   69   &   2s$^2$2p$^5$4d	  &  	  $^3$P$^o_{2}$    &   209.7364    & 210.0653  & 209.5701   &  210.0884 &   &	    1.744-13   \\
   70   &   2s$^2$2p$^5$4d	  &  	  $^1$F$^o_{3}$    &   209.7832    & 210.1147  & 209.6157   &  210.1325 &   &	    1.750-13   \\
   71   &   2s$^2$2p$^5$4d	  &  	  $^3$D$^o_{1}$    &   209.9192    & 210.2458  & 209.7448   &  210.2648 &   &	    2.381-14   \\
   72   &   2s$^2$2p$^5$4f	  &  	  $^3$G$  _{3}$    &   210.8816    & 211.2201  & 210.7060   &  211.2571 &   &	    8.423-14   \\
   73   &   2s$^2$2p$^5$4f	  &  	  $^3$F$  _{2}$    &   210.9115    & 211.2487  & 210.7362   &  211.2791 &   &	    8.427-14   \\
   74   &   2s$^2$2p$^5$4f	  &  	  $^1$G$  _{4}$    &   210.9436    & 211.2829  & 210.7674   &  211.3159 &   &	    8.492-14   \\
   75   &   2s$^2$2p$^5$4f	  &  	  $^3$D$  _{3}$    &   210.9487    & 211.2869  & 210.7725   &  211.3159 &   &	    8.335-14   \\
   76   &   2s2p$^6$4s  	  &  	  $^3$S$  _{1}$    &   219.9878    & 220.2881  & 219.9299   &  220.3194 &   &	    2.957-13   \\
   77   &   2s2p$^6$4s  	  &  	  $^1$S$  _{0}$    &   220.3037    & 220.5991  & 220.2466   &  220.6134 &   &	    3.127-13   \\
   78   &   2s2p$^6$4p  	  &  	  $^3$P$^o_{0}$    &   221.8067    & 222.1002  & 221.7550   &  222.1201 &   &	    2.714-13   \\
   79   &   2s2p$^6$4p  	  &  	  $^3$P$^o_{1}$    &   221.8169    & 222.1140  & 221.7623   &  222.1495 &   &	    1.018-13   \\
   80   &   2s2p$^6$4p  	  &  	  $^3$P$^o_{2}$    &   222.2897    & 222.5686  & 222.1888   &  222.7228 &   &	    3.191-13   \\
\hline
\end{tabular}
\end{table}

\clearpage
\newpage
\setcounter{table}{3} 
\begin{table}
\caption{... continued.}
\small 
\centering
\begin{tabular}{rllrrrrrrrrr} \hline
Index  &     Configuration        & Level                  & GRASP1 & GRASP2        & FAC  &  DFS    &  YODA     &       $\tau$    (GRASP1)    \\
\hline
   81   &   2s2p$^6$4p  	  &  	  $^1$P$^o_{1}$    &   222.4682    & 222.7721  & 222.4021   &  222.7963 &   &	    5.936-14   \\
   82   &   2s$^2$2p$^5$5s	  &  	  $^1$P$^o_{1}$    &   222.6048    & 222.8365  & 222.4369   &		&   &	    2.513-13   \\
   83   &   2s$^2$2p$^5$5s	  &  	  $^3$P$^o_{2}$    &   222.6489    & 222.9041  & 222.5133   &		&   &	    4.010-13   \\
   84   &   2s$^2$2p$^5$5p	  &  	  $^3$S$  _{1}$    &   223.4433    & 223.6654  & 223.2605   &		&   &	    3.891-13   \\
   85   &   2s$^2$2p$^5$5p	  &  	  $^3$D$  _{2}$    &   223.4497    & 223.6750  & 223.2688   &		&   &	    3.711-13   \\
   86   &   2s$^2$2p$^5$5p	  &  	  $^3$D$  _{3}$    &   223.7106    & 223.9432  & 223.5300   &		&   &	    4.028-13   \\
   87   &   2s$^2$2p$^5$5p	  &  	  $^1$P$  _{1}$    &   223.7394    & 223.9689  & 223.5557   &		&   &	    4.095-13   \\
   88   &   2s$^2$2p$^5$5p	  &  	  $^3$P$  _{2}$    &   223.7892    & 224.0166  & 223.6064   &		&   &	    4.206-13   \\
   89   &   2s$^2$2p$^5$5p	  &  	  $^1$S$  _{0}$    &   224.0789    & 224.2965  & 223.8943   &		&   &	    4.332-13   \\
   90   &   2s2p$^6$4d  	  &  	  $^3$D$  _{1}$    &   224.4288    & 224.7380  & 224.3693   &  224.7881 &   &	    1.733-13   \\
   91   &   2s2p$^6$4d  	  &  	  $^3$D$  _{2}$    &   224.4779    & 224.7928  & 224.4184   &  224.8249 &   &	    1.720-13   \\
   92   &   2s2p$^6$4d  	  &  	  $^3$D$  _{3}$    &   224.5842    & 224.9046  & 224.5246   &  224.9278 &   &	    1.703-13   \\
   93   &   2s$^2$2p$^5$5d	  &  	  $^3$P$^o_{0}$    &   224.6993    & 224.9276  & 224.5216   &		&   &	    2.400-13   \\
   94   &   2s$^2$2p$^5$5d	  &  	  $^3$P$^o_{1}$    &   224.7374    & 224.9689  & 224.5588   &		&   &	    2.383-13   \\
   95   &   2s$^2$2p$^5$5d	  &  	  $^3$F$^o_{3}$    &   224.7747    & 225.0091  & 224.5944   &		&   &	    2.405-13   \\
   96   &   2s$^2$2p$^5$5d	  &  	  $^3$D$^o_{2}$    &   224.7997    & 225.0306  & 224.6191   &		&   &	    2.418-13   \\
   97   &   2s$^2$2p$^5$5d	  &  	  $^3$F$^o_{4}$    &   224.8111    & 225.0491  & 224.6317   &		&   &	    2.401-13   \\
   98   &   2s$^2$2p$^5$5d	  &  	  $^1$D$^o_{2}$    &   224.8502    & 225.0839  & 224.6686   &		&   &	    2.407-13   \\
   99   &   2s2p$^6$4d  	  &  	  $^1$D$  _{2}$    &   224.7977    & 225.1145  & 224.7342   &  225.1336 &   &	    1.696-13   \\
  100   &   2s$^2$2p$^5$5d	  &  	  $^3$D$^o_{3}$    &   224.8972    & 225.1308  & 224.7150   &		&   &	    2.409-13   \\
  101   &   2s$^2$2p$^5$5d	  &  	  $^1$P$^o_{1}$    &   225.0872    & 225.3180  & 224.8959   &		&   &	    3.097-14   \\
  102   &   2s$^2$2p$^5$5g	  &  	  $^3$F$^o_{2}$    &   225.3178    & 225.5788  & 225.1566   &		&   &	    2.050-13   \\
  103   &   2s$^2$2p$^5$5g	  &  	  $^3$F$^o_{3}$    &   225.3502    & 225.6094  & 225.1846   &		&   &	    2.157-13   \\
  104   &   2s$^2$2p$^5$5f	  &  	  $^3$G$  _{4}$    &   225.3857    & 225.6287  & 225.2042   &		&   &	    1.608-13   \\
  105   &   2s$^2$2p$^5$5f	  &  	  $^3$D$  _{1}$    &   225.3960    & 225.6370  & 225.2177   &		&   &	    1.486-13   \\
  106   &   2s$^2$2p$^5$5f	  &  	  $^3$G$  _{5}$    &   225.4076    & 225.6509  & 225.2261   &		&   &	    1.611-13   \\
  107   &   2s$^2$2p$^5$5f	  &  	  $^3$D$  _{2}$    &   225.4161    & 225.6579  & 225.2372   &		&   &	    1.515-13   \\
  108   &   2s$^2$2p$^5$5f	  &  	  $^3$F$  _{3}$    &   225.4264    & 225.6673  & 225.2451   &		&   &	    1.586-13   \\
  109   &   2s$^2$2p$^5$5g	  &  	  $^3$G$^o_{3}$    &   225.4225    & 225.6782  & 225.2485   &		&   &	    2.272-13   \\
  110   &   2s$^2$2p$^5$5f	  &  	  $^1$D$  _{2}$    &   225.4457    & 225.6869  & 225.2651   &		&   &	    1.602-13   \\
  111   &   2s$^2$2p$^5$5g	  &  	  $^3$G$^o_{4}$    &   225.4377    & 225.6930  & 225.2631   &		&   &	    2.281-13   \\
  112   &   2s$^2$2p$^5$5f	  &  	  $^1$F$  _{3}$    &   225.4515    & 225.6938  & 225.2713   &		&   &	    1.551-13   \\
  113   &   2s$^2$2p$^5$5f	  &  	  $^3$F$  _{4}$    &   225.4566    & 225.6982  & 225.2750   &		&   &	    1.603-13   \\
  114   &   2s$^2$2p$^5$5g	  &  	  $^1$H$^o_{5}$    &   225.4899    & 225.7330  & 225.2931   &		&   &	    2.882-13   \\
  115   &   2s$^2$2p$^5$5g	  &  	  $^3$H$^o_{6}$    &   225.5069    & 225.7498  & 225.3100   &		&   &	    2.889-13   \\
  116   &   2s$^2$2p$^5$5g	  &  	  $^1$G$^o_{4}$    &   225.5151    & 225.7571  & 225.3186   &		&   &	    2.882-13   \\
  117   &   2s$^2$2p$^5$5g	  &  	  $^3$G$^o_{5}$    &   225.5310    & 225.7730  & 225.3345   &		&   &	    2.892-13   \\
  118   &   2s2p$^6$4f  	  &  	  $^3$F$^o_{3}$    &   225.7962    & 226.1104  & 225.7044   &  226.0597 &   &	    8.881-14   \\
  119   &   2s2p$^6$4f  	  &  	  $^3$F$^o_{4}$    &   225.8311    & 226.1473  & 225.7417   &  226.1111 &   &	    8.798-14   \\
  120   &   2s2p$^6$4f            &       $^3$F$^o_{2}$    &   225.8669    & 226.1719  & 225.7639   &  226.0523 &   &       9.089-14   \\
\hline
\end{tabular}
\end{table}

\clearpage
\newpage
\setcounter{table}{3} 
\begin{table}
\caption{... continued.}
\small 
\centering
\begin{tabular}{rllrrrrrrrrr} \hline
Index  &     Configuration        & Level                  & GRASP1 & GRASP2        & FAC  &  DFS    &  YODA     &       $\tau$    (GRASP1)    \\
\hline
  121   &   2s2p$^6$4f  	  &  	  $^1$F$^o_{3}$   &    225.9203   & 226.2297  & 225.8217   &  226.1332 &   &	   8.967-14   \\
  122   &   2s$^2$2p$^5$5s	  &  	  $^3$P$^o_{0}$   &    228.3506   & 228.6719  & 228.1423   &	       &   &	   4.345-13   \\
  123   &   2s$^2$2p$^5$5s	  &  	  $^3$P$^o_{1}$   &    228.3696   & 228.6922  & 228.1616   &	       &   &	   3.699-13   \\
  124   &   2s$^2$2p$^5$5p	  &  	  $^3$D$  _{1}$   &    229.2465   & 229.5683  & 229.2483   &	       &   &	   3.776-13   \\
  125   &   2s$^2$2p$^5$5p	  &  	  $^3$P$  _{0}$   &    229.4556   & 229.7686  & 229.0393   &	       &   &	   3.932-13   \\
  126   &   2s$^2$2p$^5$5p	  &  	  $^3$P$  _{1}$   &    229.5449   & 229.8690  & 229.3380   &	       &   &	   4.238-13   \\
  127   &   2s$^2$2p$^5$5p	  &  	  $^1$D$  _{2}$   &    229.5601   & 229.8870  & 229.3535   &	       &   &	   4.207-13   \\
  128   &   2s$^2$2p$^5$5d	  &  	  $^3$F$^o_{2}$   &    230.5663   & 230.8966  & 230.3626   &	       &   &	   2.415-13   \\
  129   &   2s$^2$2p$^5$5d	  &  	  $^3$P$^o_{2}$   &    230.6384   & 230.9698  & 230.4346   &	       &   &	   2.397-13   \\
  130   &   2s$^2$2p$^5$5d	  &  	  $^1$F$^o_{3}$   &    230.6603   & 230.9929  & 230.4558   &	       &   &	   2.407-13   \\
  131   &   2s$^2$2p$^5$5d	  &  	  $^3$D$^o_{1}$   &    230.7145   & 231.0425  & 230.5047   &	       &   &	   4.944-14   \\
  132   &   2s$^2$2p$^5$5f	  &  	  $^3$G$  _{3}$   &    231.2020   & 231.5406  & 230.9968   &	       &   &	   1.599-13   \\
  133   &   2s$^2$2p$^5$5f	  &  	  $^3$F$  _{2}$   &    231.2205   & 231.5584  & 231.0153   &	       &   &	   1.598-13   \\
  134   &   2s$^2$2p$^5$5f	  &  	  $^1$G$  _{4}$   &    231.2348   & 231.5740  & 231.0291   &	       &   &	   1.612-13   \\
  135   &   2s$^2$2p$^5$5f	  &  	  $^3$D$  _{3}$   &    231.2366   & 231.5751  & 231.0308   &	       &   &	   1.576-13   \\
  136   &   2s$^2$2p$^5$5g	  &  	  $^3$H$^o_{4}$   &    231.3059   & 231.6452  & 231.0853   &	       &   &	   2.883-13   \\
  137   &   2s$^2$2p$^5$5g	  &  	  $^1$F$^o_{3}$   &    231.3132   & 231.6521  & 231.0928   &	       &   &	   2.839-13   \\
  138   &   2s$^2$2p$^5$5g	  &  	  $^3$H$^o_{5}$   &    231.3233   & 231.6626  & 231.1026   &	       &   &	   2.891-13   \\
  139   &   2s$^2$2p$^5$5g	  &  	  $^3$F$^o_{4}$   &    231.3301   & 231.6691  & 231.1097   &	       &   &	   2.844-13   \\
\hline				  				        			
\end{tabular}										            			    
\begin {flushleft}									            			    
\begin{tabbing} 									      	
aaaaaaaaaaaaaaaaaaaaaaaaaaaaaaaaaaaa\= \kill						      	       
GRASP1: Present results  with the {\sc grasp} code  for 3948   level calculations {\em including} Breit and QED effects \\		
GRASP2: Present results  with the {\sc grasp} code for 3948   level calculations {\em excluding} Breit and QED effects \\		
FAC: Present results  with the {\sc fac} code   for  93~437  level calculations \\ 
DFS: Earlier results of Zhang and Sampson \cite{zs2}  \\
YODA: Earlier results of Hagelstein and Jung\cite{hag}  \\			 			      		      	
\end{tabbing}										      	
\end {flushleft}									      	
\end{table}

Assessing the accuracy of our calculated results for $A$-values (and other related parameters) is not straightforward. This is because no measurements are available for any transition of the ions concerned. However, limited theoretical results are available in the literature which will perhaps be helpful for some accuracy assessments. For Sc~XIII,  J\"{o}nsson et al. \cite{jag} have listed $A$-values for  the 1--3 E1 (4.635$\times$10$^{10}$~s$^{-1}$), 2--3 E1 (1.968$\times$10$^{10}$~s$^{-1}$), 1--2 M1 (9.773$\times$10$^{2}$~s$^{-1}$), and 1--2 E2 (3.849$\times$10$^{-2}$~s$^{-1}$) transitions, which match very well (within 10\%) with our corresponding results of 5.143$\times$10$^{10}$, 2.193$\times$10$^{10}$, 9.571$\times$10$^{2}$, and 3.849$\times$10$^{-2}$~s$^{-1}$, respectively. However, this direct comparison of $A$-values is very limited. Some further assessments of accuracy can be made by comparing the length and velocity forms (i.e. the Babushkin and Coulomb gauges in the relativistic terms) of the $A$-values, and their ratio (R)  for all E1 transitions are listed in Tables~6--8. Ideally R should be closer to unity but in practice it is not, particularly for the weak(er) transitions. For many strong transitions with $f$ $\ge$ 0.1, R is within 10\% of unity as may be noted for the 1--49/52/55/56/59 transitions in Table~6. However, for 22 transitions (all with $f$ $\le$ 0.2) R is up to 2 and examples include 9--26/78, 10--25/79 and 25--50/89. Similarly,  for a few very weak transitions R can be  up to several orders of magnitude, and some examples are:  3--18 ($f$ = 5.9$\times$10$^{-6}$, R = 164), 3--21 ($f$ = 3.8$\times$10$^{-4}$, R = 83) and 3--23 ($f$ = 4.0$\times$10$^{-5}$, R = 357). For such weak transitions the modelling of plasmas is not affected and similar large values of R are often noted for almost all ions in any large calculation. 
 
\begin{table}
\caption{Comparison of energies (in Ryd)  for the lowest 37 levels of  Y XXX.}
\small 
\centering
\begin{tabular}{rllrrrrrrrr} \hline
Index  &     Configuration        & Level                  & FAC1          & FAC2      & FAC3     & GRASP1    & GRASP2       & GRASP3      \\
 \hline
    1   &   2s$^2$2p$^6$	  &  	  $^1$S$  _{0}$    &    0.0000     &   0.0000  &   0.0000 &   0.0000  &      0.0000  &     0.0000  \\
    2   &   2s$^2$2p$^5$3s	  &  	  $^3$P$^o_{2}$    &  146.6164     & 146.5697  & 146.5497 & 146.7014  &  	     &   146.6987  \\
    3   &   2s$^2$2p$^5$3s	  &  	  $^1$P$^o_{1}$    &  146.8884     & 146.8272  & 146.8078 & 146.9787  &  	     &   146.9606  \\
    4   &   2s$^2$2p$^5$3p	  &  	  $^3$S$  _{1}$    &  150.9125     & 150.8425  & 150.8223 & 150.9653  &    151.0112  &   150.9658  \\
    5   &   2s$^2$2p$^5$3p	  &  	  $^3$D$  _{2}$    &  151.1260     & 151.0538  & 151.0345 & 151.1747  &  	     &   151.1804  \\
    6   &   2s$^2$2p$^5$3p	  &  	  $^3$D$  _{3}$    &  152.4161     & 152.3427  & 152.3230 & 152.4651  &    152.5129  &   152.4685  \\
    7   &   2s$^2$2p$^5$3p	  &  	  $^1$P$  _{1}$    &  152.4397     & 152.3634  & 152.3437 & 152.4908  &    152.5414  &   152.4932  \\
    8   &   2s$^2$2p$^5$3s	  &  	  $^3$P$^o_{0}$    &  152.3800     & 152.3298  & 152.3099 & 152.4813  &    152.5560  &   152.4791  \\
    9   &   2s$^2$2p$^5$3s	  &  	  $^3$P$^o_{1}$    &  152.5137     & 152.4558  & 152.4363 & 152.6247  &  	     &   152.6077  \\
   10   &   2s$^2$2p$^5$3p	  &  	  $^3$P$  _{2}$    &  152.8034     & 152.7307  & 152.7116 & 152.8575  &  	     &   152.8584  \\
   11   &   2s$^2$2p$^5$3p	  &  	  $^3$P$  _{0}$    &  154.2124     & 154.1432  & 154.1154 & 154.2961  &    154.3980  &   154.2837  \\
   12   &   2s$^2$2p$^5$3p	  &  	  $^3$D$  _{1}$    &  156.7825     & 156.7041  & 156.6846 & 156.8470  &    156.9004  &   156.8517  \\
   13   &   2s$^2$2p$^5$3d	  &  	  $^3$P$^o_{0}$    &  157.5915     & 157.5213  & 157.5000 & 157.6708  &    157.7660  &   157.6799  \\
   14   &   2s$^2$2p$^5$3d	  &  	  $^3$P$^o_{1}$    &  157.8001     & 157.7287  & 157.7075 & 157.8813  &  	     &   157.8890  \\
   15   &   2s$^2$2p$^5$3d	  &  	  $^3$F$^o_{3}$    &  158.0818     & 158.0112  & 157.9895 & 158.1763  &    158.2056  &   158.1738  \\
   16   &   2s$^2$2p$^5$3d	  &  	  $^3$D$^o_{2}$    &  158.1807     & 158.1077  & 158.0866 & 158.2663  &    158.2950  &   158.2712  \\
   17   &   2s$^2$2p$^5$3d	  &  	  $^3$F$^o_{4}$    &  158.2009     & 158.1363  & 158.1143 & 158.2879  &    158.3356  &   158.2876  \\
   18   &   2s$^2$2p$^5$3p	  &  	  $^3$P$  _{1}$    &  158.2446     & 158.1690  & 158.1492 & 158.3102  &    158.3604  &   158.3146  \\
   19   &   2s$^2$2p$^5$3p	  &  	  $^1$D$  _{2}$    &  158.3547     & 158.2770  & 158.2578 & 158.4209  &  	     &   158.4251  \\
   20   &   2s$^2$2p$^5$3d	  &  	  $^1$D$^o_{2}$    &  158.4372     & 158.3602  & 158.3391 & 158.5266  &    158.5801  &   158.5283  \\
   21   &   2s$^2$2p$^5$3p	  &  	  $^1$S$  _{0}$    &  158.5336     & 158.4648  & 158.4291 & 158.5319  &    158.5608  &   158.6252  \\
   22   &   2s$^2$2p$^5$3d	  &  	  $^3$D$^o_{3}$    &  158.7122     & 158.6354  & 158.6145 & 158.8041  &    158.8350  &   158.8049  \\
   23   &   2s$^2$2p$^5$3d	  &  	  $^3$D$^o_{1}$    &  159.6610     & 159.5862  & 159.5589 & 159.7764  &  	     &   159.7700  \\
   24   &   2s$^2$2p$^5$3d	  &  	  $^3$F$^o_{2}$    &  163.8100     & 163.7367  & 163.7151 & 163.9070  &    163.9513  &   163.9104  \\
   25   &   2s$^2$2p$^5$3d	  &  	  $^3$P$^o_{2}$    &  164.1176     & 164.0390  & 164.0182 & 164.2192  &    164.2507  &   164.2275  \\
   26   &   2s$^2$2p$^5$3d	  &  	  $^1$F$^o_{3}$    &  164.2434     & 164.1664  & 164.1453 & 164.3523  &    164.3840  &   164.3548  \\
   27   &   2s$^2$2p$^5$3d	  &  	  $^1$P$^o_{1}$    &  164.7694     & 164.6913  & 164.6642 & 164.8795  &  	     &   164.8886  \\
   28   &   2s2p$^6$3s  	  &  	  $^3$S$  _{1}$    &  167.6548     & 167.5336  & 167.5051 & 	      &  	     &   167.5933  \\
   29   &   2s2p$^6$3s  	  &  	  $^1$S$  _{0}$    &  168.5527     & 168.3882  & 168.3525 & 	      &  	     &   168.4649  \\
   30   &   2s2p$^6$3p  	  &  	  $^3$P$^o_{0}$    &  172.0201     & 171.9109  & 171.8826 & 	      &  	     &   171.9546  \\
   31   &   2s2p$^6$3p  	  &  	  $^3$P$^o_{1}$    &  172.1062     & 171.9936  & 171.9655 & 	      &  	     &   172.0393  \\
   32   &   2s2p$^6$3p  	  &  	  $^3$P$^o_{2}$    &  173.4597     & 173.3526  & 173.3240 & 	      &  	     &   173.3963  \\
   33   &   2s2p$^6$3p  	  &  	  $^1$P$^o_{1}$    &  173.6801     & 173.5627  & 173.5349 & 	      &  	     &   173.6114  \\
   34   &   2s2p$^6$3d  	  &  	  $^3$D$  _{1}$    &  178.8518     & 178.7405  & 178.7101 & 	      &  	     &   178.8170  \\
   35   &   2s2p$^6$3d  	  &  	  $^3$D$  _{2}$    &  178.9334     & 178.8218  & 178.7914 & 	      &  	     &   178.8986  \\
   36   &   2s2p$^6$3d  	  &  	  $^3$D$  _{3}$    &  179.1391     & 179.0286  & 178.9983 & 	      &  	     &   179.1053  \\
   37   &   2s2p$^6$3d  	  &  	  $^1$D$  _{2}$    &  179.7807     & 179.6561  & 179.6257 & 	      &  	     &   179.7476  \\
\hline				  				        			
\end{tabular}										             				    
\begin {flushleft}									            			    
\begin{tabbing} 									      	
aaaaaaaaaaaaaaaaaaaaaaaaaaaaaaaaaaaa\= \kill						      	       
FAC1: Present results  with the {\sc fac} code for 3948   level calculations \\
FAC2: Present results  with the {\sc fac} code for 17~729 level calculations  \\		
FAC3: Present results  with the {\sc fac} code for  93~437 level calculations  \\ 
GRASP1: Earlier results of Cogordan and Lunell  \cite{cl} with the {\sc grasp} code \\
GRASP2: Earlier results of Quinet et al. \cite{pq} with the {\sc grasp} code \\	
GRASP3: Present results  with the {\sc grasp} code for 3948   level calculations \\	 			      		      	
\end{tabbing}										      	
\end {flushleft}					
\end{table}

\setcounter{table}{5} 
\begin{table}
\caption{Transition wavelengths ($\lambda_{ij}$ in \AA), radiative rates (A$_{ji}$ in s$^{-1}$), oscillator strengths (f$_{ij}$, dimensionless), and line     
strengths (S, in atomic units) for electric dipole (E1), and A$_{ji}$ for E2, M1 and M2 transitions in Sc~XIII. The last column gives the ratio R  of  the velocity and length forms of A(E1). $a{\pm}b \equiv a{\times}$10$^{{\pm}b}$.}
\small 
\centering
\begin{tabular}{rrrrrrrrrr} \hline        
$i$ & $j$ & $\lambda_{ij}$ & A$^{{\rm E1}}_{ji}$  & f$^{{\rm E1}}_{ij}$ & S$^{{\rm E1}}$ & A$^{{\rm E2}}_{ji}$  & A$^{{\rm M1}}_{ij}$ & A$^{{\rm M2}}$ & R  \\
\hline                                    
    1 &    2 &  2.656$+$03 &  0.000$+$00 &  0.000$+$00 &  0.000$+$00 &  3.849$-$02 &  9.570$+$02 &  0.000$+$00 &   0.0$+$00 \\
    1 &    3 &  1.284$+$02 &  5.143$+$10 &  6.354$-$02 &  1.074$-$01 &  0.000$+$00 &  0.000$+$00 &  4.272$+$02 &   8.4$-$01 \\
    1 &    4 &  2.856$+$01 &  1.016$+$10 &  1.863$-$03 &  7.009$-$04 &  0.000$+$00 &  0.000$+$00 &  4.296$+$04 &   9.3$-$01 \\
    1 &    5 &  2.842$+$01 &  2.717$+$11 &  3.290$-$02 &  1.231$-$02 &  0.000$+$00 &  0.000$+$00 &  7.041$+$03 &   9.4$-$01 \\
    1 &    6 &  2.829$+$01 &  1.706$+$09 &  1.023$-$04 &  3.812$-$05 &  0.000$+$00 &  0.000$+$00 &  1.414$+$04 &   1.1$+$00 \\
    1 &    7 &  2.822$+$01 &  5.591$+$11 &  6.676$-$02 &  2.481$-$02 &  0.000$+$00 &  0.000$+$00 &  1.737$+$03 &   9.4$-$01 \\
    1 &    8 &  2.807$+$01 &  3.941$+$11 &  2.327$-$02 &  8.601$-$03 &  0.000$+$00 &  0.000$+$00 &  1.264$+$04 &   9.5$-$01 \\
    1 &    9 &  2.770$+$01 &  3.539$+$11 &  6.107$-$02 &  2.228$-$02 &  0.000$+$00 &  0.000$+$00 &  1.063$+$04 &   9.3$-$01 \\
    1 &   10 &  2.769$+$01 &  1.897$+$09 &  2.181$-$04 &  7.952$-$05 &  0.000$+$00 &  0.000$+$00 &  3.051$+$03 &   8.7$-$01 \\
    1 &   11 &  2.721$+$01 &  0.000$+$00 &  0.000$+$00 &  0.000$+$00 &  1.377$+$06 &  1.238$+$03 &  0.000$+$00 &   0.0$+$00 \\
    1 &   12 &  2.720$+$01 &  0.000$+$00 &  0.000$+$00 &  0.000$+$00 &  1.759$+$07 &  1.803$+$04 &  0.000$+$00 &   0.0$+$00 \\
    1 &   13 &  2.710$+$01 &  0.000$+$00 &  0.000$+$00 &  0.000$+$00 &  3.160$+$07 &  9.887$+$03 &  0.000$+$00 &   0.0$+$00 \\
    1 &   14 &  2.705$+$01 &  0.000$+$00 &  0.000$+$00 &  0.000$+$00 &  7.016$+$06 &  0.000$+$00 &  0.000$+$00 &   0.0$+$00 \\
    1 &   15 &  2.701$+$01 &  0.000$+$00 &  0.000$+$00 &  0.000$+$00 &  1.986$+$08 &  2.216$+$03 &  0.000$+$00 &   0.0$+$00 \\
    1 &   16 &  2.700$+$01 &  1.483$+$11 &  8.105$-$03 &  2.882$-$03 &  0.000$+$00 &  0.000$+$00 &  3.201$+$04 &   1.0$+$00 \\
    1 &   17 &  2.690$+$01 &  0.000$+$00 &  0.000$+$00 &  0.000$+$00 &  1.302$+$08 &  5.680$+$03 &  0.000$+$00 &   0.0$+$00 \\
    1 &   18 &  2.689$+$01 &  0.000$+$00 &  0.000$+$00 &  0.000$+$00 &  2.984$+$07 &  2.450$+$01 &  0.000$+$00 &   0.0$+$00 \\
    1 &   19 &  2.686$+$01 &  0.000$+$00 &  0.000$+$00 &  0.000$+$00 &  1.118$+$04 &  2.589$+$03 &  0.000$+$00 &   0.0$+$00 \\
    1 &   20 &  2.682$+$01 &  0.000$+$00 &  0.000$+$00 &  0.000$+$00 &  1.456$+$08 &  6.430$+$03 &  0.000$+$00 &   0.0$+$00 \\
    1 &   21 &  2.680$+$01 &  0.000$+$00 &  0.000$+$00 &  0.000$+$00 &  2.498$+$08 &  2.819$+$00 &  0.000$+$00 &   0.0$+$00 \\
    1 &   22 &  2.673$+$01 &  0.000$+$00 &  0.000$+$00 &  0.000$+$00 &  8.681$+$06 &  3.929$+$03 &  0.000$+$00 &   0.0$+$00 \\
    1 &   23 &  2.669$+$01 &  0.000$+$00 &  0.000$+$00 &  0.000$+$00 &  7.119$+$07 &  2.847$+$03 &  0.000$+$00 &   0.0$+$00 \\
    1 &   24 &  2.668$+$01 &  0.000$+$00 &  0.000$+$00 &  0.000$+$00 &  5.727$+$07 &  6.973$+$03 &  0.000$+$00 &   0.0$+$00 \\
    1 &   25 &  2.640$+$01 &  0.000$+$00 &  0.000$+$00 &  0.000$+$00 &  1.702$+$07 &  2.971$+$02 &  0.000$+$00 &   0.0$+$00 \\
    1 &   26 &  2.635$+$01 &  0.000$+$00 &  0.000$+$00 &  0.000$+$00 &  1.932$+$08 &  0.000$+$00 &  0.000$+$00 &   0.0$+$00 \\
    1 &   27 &  2.623$+$01 &  0.000$+$00 &  0.000$+$00 &  0.000$+$00 &  1.504$+$07 &  9.730$+$03 &  0.000$+$00 &   0.0$+$00 \\
    1 &   28 &  2.620$+$01 &  0.000$+$00 &  0.000$+$00 &  0.000$+$00 &  5.473$+$07 &  9.653$+$03 &  0.000$+$00 &   0.0$+$00 \\
    1 &   29 &  2.588$+$01 &  0.000$+$00 &  0.000$+$00 &  0.000$+$00 &  2.103$+$07 &  4.039$+$02 &  0.000$+$00 &   0.0$+$00 \\
    1 &   30 &  2.585$+$01 &  0.000$+$00 &  0.000$+$00 &  0.000$+$00 &  1.236$+$08 &  1.065$+$04 &  0.000$+$00 &   0.0$+$00 \\
    1 &   31 &  2.564$+$01 &  0.000$+$00 &  0.000$+$00 &  0.000$+$00 &  1.784$+$07 &  1.429$+$03 &  0.000$+$00 &   0.0$+$00 \\
    1 &   32 &  2.556$+$01 &  0.000$+$00 &  0.000$+$00 &  0.000$+$00 &  2.357$+$05 &  8.866$+$02 &  0.000$+$00 &   0.0$+$00 \\
    1 &   33 &  2.555$+$01 &  0.000$+$00 &  0.000$+$00 &  0.000$+$00 &  0.000$+$00 &  0.000$+$00 &  7.415$+$05 &   0.0$+$00 \\
    1 &   34 &  2.554$+$01 &  4.054$+$07 &  5.948$-$06 &  2.001$-$06 &  0.000$+$00 &  0.000$+$00 &  4.591$+$03 &   6.6$-$01 \\
    1 &   35 &  2.552$+$01 &  2.277$+$09 &  2.223$-$04 &  7.468$-$05 &  0.000$+$00 &  0.000$+$00 &  1.059$+$04 &   9.6$-$01 \\
    1 &   36 &  2.548$+$01 &  9.740$+$08 &  4.741$-$05 &  1.591$-$05 &  0.000$+$00 &  0.000$+$00 &  1.489$+$05 &   1.0$+$00 \\
    1 &   38 &  2.536$+$01 &  0.000$+$00 &  0.000$+$00 &  0.000$+$00 &  0.000$+$00 &  0.000$+$00 &  1.461$+$05 &   0.0$+$00 \\
    1 &   39 &  2.531$+$01 &  2.690$+$11 &  1.292$-$02 &  4.304$-$03 &  0.000$+$00 &  0.000$+$00 &  2.884$+$05 &   9.1$-$01 \\
    1 &   40 &  2.526$+$01 &  3.929$+$11 &  5.638$-$02 &  1.876$-$02 &  0.000$+$00 &  0.000$+$00 &  3.032$+$04 &   9.5$-$01 \\
\hline  
\end{tabular}
\end{table}

\clearpage
\newpage
\setcounter{table}{5} 
\begin{table}
\caption{... continued.}
\small 
\centering
\begin{tabular}{rrrrrrrrrr} \hline        
$i$ & $j$ & $\lambda_{ij}$ & A$^{{\rm E1}}_{ji}$  & f$^{{\rm E1}}_{ij}$ & S$^{{\rm E1}}$ & A$^{{\rm E2}}_{ji}$  & A$^{{\rm M1}}_{ij}$ & A$^{{\rm M2}}$ & R  \\
\hline 
    1 &   41 &  2.526$+$01 &  3.991$+$11 &  3.818$-$02 &  1.270$-$02 &  0.000$+$00 &  0.000$+$00 &  8.862$+$04 &   9.3$-$01 \\
    1 &   42 &  2.522$+$01 &  2.258$+$11 &  1.077$-$02 &  3.576$-$03 &  0.000$+$00 &  0.000$+$00 &  8.113$+$05 &   9.3$-$01 \\
    1 &   43 &  2.521$+$01 &  2.046$+$10 &  1.950$-$03 &  6.474$-$04 &  0.000$+$00 &  0.000$+$00 &  8.295$+$02 &   9.5$-$01 \\
    1 &   44 &  2.522$+$01 &  0.000$+$00 &  0.000$+$00 &  0.000$+$00 &  0.000$+$00 &  0.000$+$00 &  3.590$+$03 &   0.0$+$00 \\
    1 &   45 &  2.520$+$01 &  2.956$+$11 &  4.221$-$02 &  1.400$-$02 &  0.000$+$00 &  0.000$+$00 &  3.244$+$05 &   9.5$-$01 \\
    1 &   46 &  2.516$+$01 &  7.467$+$11 &  7.084$-$02 &  2.347$-$02 &  0.000$+$00 &  0.000$+$00 &  1.225$+$05 &   9.4$-$01 \\
    1 &   47 &  2.514$+$01 &  2.270$+$11 &  3.226$-$02 &  1.068$-$02 &  0.000$+$00 &  0.000$+$00 &  2.500$+$03 &   9.5$-$01 \\
    1 &   48 &  2.505$+$01 &  7.205$+$10 &  6.781$-$03 &  2.237$-$03 &  0.000$+$00 &  0.000$+$00 &  9.144$+$04 &   9.3$-$01 \\
    1 &   49 &  2.502$+$01 &  2.278$+$12 &  3.207$-$01 &  1.057$-$01 &  0.000$+$00 &  0.000$+$00 &  4.764$+$03 &   9.5$-$01 \\
    1 &   50 &  2.483$+$01 &  0.000$+$00 &  0.000$+$00 &  0.000$+$00 &  0.000$+$00 &  0.000$+$00 &  9.072$+$03 &   0.0$+$00 \\
    1 &   52 &  2.470$+$01 &  4.834$+$12 &  2.212$-$01 &  7.195$-$02 &  0.000$+$00 &  0.000$+$00 &  3.265$+$05 &   9.0$-$01 \\
    1 &   53 &  2.469$+$01 &  3.016$+$11 &  4.134$-$02 &  1.344$-$02 &  0.000$+$00 &  0.000$+$00 &  6.072$+$04 &   9.6$-$01 \\
    1 &   54 &  2.467$+$01 &  0.000$+$00 &  0.000$+$00 &  0.000$+$00 &  0.000$+$00 &  0.000$+$00 &  3.246$+$04 &   0.0$+$00 \\
    1 &   55 &  2.460$+$01 &  6.680$+$12 &  6.059$-$01 &  1.963$-$01 &  0.000$+$00 &  0.000$+$00 &  2.424$+$05 &   9.2$-$01 \\
    1 &   56 &  2.458$+$01 &  6.076$+$12 &  8.259$-$01 &  2.674$-$01 &  0.000$+$00 &  0.000$+$00 &  1.651$+$04 &   9.6$-$01 \\
    1 &   57 &  2.451$+$01 &  7.517$+$11 &  6.772$-$02 &  2.186$-$02 &  0.000$+$00 &  0.000$+$00 &  7.996$+$04 &   9.6$-$01 \\
    1 &   58 &  2.451$+$01 &  1.608$+$12 &  7.245$-$02 &  2.339$-$02 &  0.000$+$00 &  0.000$+$00 &  9.940$+$04 &   9.2$-$01 \\
    1 &   59 &  2.417$+$01 &  7.732$+$11 &  1.016$-$01 &  3.233$-$02 &  0.000$+$00 &  0.000$+$00 &  1.510$+$05 &   9.6$-$01 \\
    1 &   60 &  2.413$+$01 &  6.649$+$10 &  5.804$-$03 &  1.844$-$03 &  0.000$+$00 &  0.000$+$00 &  1.598$+$02 &   1.0$+$00 \\ 
    1 &   61 &  2.387$+$01 &  0.000$+$00 &  0.000$+$00 &  0.000$+$00 &  2.093$+$02 &  2.304$+$03 &  0.000$+$00 &   0.0$+$00 \\
    1 &   62 &  2.377$+$01 &  0.000$+$00 &  0.000$+$00 &  0.000$+$00 &  6.929$+$04 &  6.434$+$02 &  0.000$+$00 &   0.0$+$00 \\
    1 &   63 &  2.368$+$01 &  0.000$+$00 &  0.000$+$00 &  0.000$+$00 &  5.313$+$04 &  1.568$+$02 &  0.000$+$00 &   0.0$+$00 \\
    1 &   64 &  2.356$+$01 &  0.000$+$00 &  0.000$+$00 &  0.000$+$00 &  1.346$+$06 &  6.397$+$01 &  0.000$+$00 &   0.0$+$00 \\
    1 &   65 &  2.344$+$01 &  0.000$+$00 &  0.000$+$00 &  0.000$+$00 &  2.864$+$06 &  2.384$+$02 &  0.000$+$00 &   0.0$+$00 \\
    1 &   66 &  2.299$+$01 &  6.378$+$07 &  5.052$-$06 &  1.529$-$06 &  0.000$+$00 &  0.000$+$00 &  3.535$+$04 &   5.4$-$01 \\
    1 &   67 &  2.285$+$01 &  0.000$+$00 &  0.000$+$00 &  0.000$+$00 &  0.000$+$00 &  0.000$+$00 &  1.067$+$05 &   0.0$+$00 \\
    1 &   68 &  2.285$+$01 &  1.207$+$11 &  1.417$-$02 &  4.262$-$03 &  0.000$+$00 &  0.000$+$00 &  7.639$+$01 &   1.0$+$00 \\
    1 &   69 &  2.280$+$01 &  1.445$+$11 &  1.126$-$02 &  3.379$-$03 &  0.000$+$00 &  0.000$+$00 &  1.961$+$04 &   9.5$-$01 \\
    1 &   70 &  2.274$+$01 &  4.807$+$11 &  5.591$-$02 &  1.675$-$02 &  0.000$+$00 &  0.000$+$00 &  7.822$+$04 &   1.0$+$00 \\
    1 &   71 &  2.274$+$01 &  2.117$+$10 &  8.206$-$04 &  2.457$-$04 &  0.000$+$00 &  0.000$+$00 &  8.406$+$01 &   9.3$-$01 \\
    1 &   72 &  2.268$+$01 &  7.750$+$11 &  5.979$-$02 &  1.786$-$02 &  0.000$+$00 &  0.000$+$00 &  1.618$+$04 &   9.5$-$01 \\
    1 &   73 &  2.266$+$01 &  3.728$+$11 &  4.306$-$02 &  1.285$-$02 &  0.000$+$00 &  0.000$+$00 &  2.096$+$04 &   1.0$+$00 \\
    1 &   74 &  2.265$+$01 &  3.029$+$10 &  1.165$-$03 &  3.473$-$04 &  0.000$+$00 &  0.000$+$00 &  2.006$+$04 &   9.8$-$01 \\
    1 &   75 &  2.265$+$01 &  2.244$+$11 &  1.726$-$02 &  5.147$-$03 &  0.000$+$00 &  0.000$+$00 &  6.875$+$03 &   9.9$-$01 \\
    1 &   76 &  2.261$+$01 &  7.117$+$11 &  2.728$-$02 &  8.124$-$03 &  0.000$+$00 &  0.000$+$00 &  8.459$+$03 &   9.6$-$01 \\
    1 &   77 &  2.256$+$01 &  1.087$+$10 &  8.295$-$04 &  2.465$-$04 &  0.000$+$00 &  0.000$+$00 &  4.842$+$03 &   7.4$-$01 \\
    1 &   78 &  2.242$+$01 &  0.000$+$00 &  0.000$+$00 &  0.000$+$00 &  3.231$+$03 &  7.871$+$02 &  0.000$+$00 &   0.0$+$00 \\
    1 &   79 &  2.241$+$01 &  0.000$+$00 &  0.000$+$00 &  0.000$+$00 &  1.042$+$05 &  1.739$+$03 &  0.000$+$00 &   0.0$+$00 \\
    1 &   80 &  2.237$+$01 &  6.381$+$11 &  2.393$-$02 &  7.047$-$03 &  0.000$+$00 &  0.000$+$00 &  5.659$+$03 &   1.0$+$00 \\
\hline  
\end{tabular}
\end{table}

\clearpage
\newpage
\setcounter{table}{5} 
\begin{table}
\caption{... continued.}
\small 
\centering
\begin{tabular}{rrrrrrrrrr} \hline        
$i$ & $j$ & $\lambda_{ij}$ & A$^{{\rm E1}}_{ji}$  & f$^{{\rm E1}}_{ij}$ & S$^{{\rm E1}}$ & A$^{{\rm E2}}_{ji}$  & A$^{{\rm M1}}_{ij}$ & A$^{{\rm M2}}$ & R  \\
\hline 
    1 &   81 &  2.180$+$01 &  0.000$+$00 &  0.000$+$00 &  0.000$+$00 &  2.445$+$04 &  2.086$+$02 &  0.000$+$00 &   0.0$+$00 \\
    1 &   82 &  2.177$+$01 &  0.000$+$00 &  0.000$+$00 &  0.000$+$00 &  8.185$+$05 &  1.952$+$01 &  0.000$+$00 &   0.0$+$00 \\
    1 &   84 &  2.174$+$01 &  0.000$+$00 &  0.000$+$00 &  0.000$+$00 &  7.207$+$06 &  1.597$+$01 &  0.000$+$00 &   0.0$+$00 \\
    1 &   85 &  2.172$+$01 &  0.000$+$00 &  0.000$+$00 &  0.000$+$00 &  4.140$+$07 &  0.000$+$00 &  0.000$+$00 &   0.0$+$00 \\
    1 &   86 &  2.167$+$01 &  0.000$+$00 &  0.000$+$00 &  0.000$+$00 &  9.654$+$07 &  1.069$+$02 &  0.000$+$00 &   0.0$+$00 \\
    1 &   87 &  2.164$+$01 &  0.000$+$00 &  0.000$+$00 &  0.000$+$00 &  5.051$+$07 &  7.469$+$01 &  0.000$+$00 &   0.0$+$00 \\
    1 &   88 &  2.161$+$01 &  0.000$+$00 &  0.000$+$00 &  0.000$+$00 &  6.636$+$08 &  0.000$+$00 &  0.000$+$00 &   0.0$+$00 \\
    1 &   89 &  2.157$+$01 &  9.573$+$10 &  6.679$-$03 &  1.898$-$03 &  0.000$+$00 &  0.000$+$00 &  3.241$+$03 &   1.0$+$00 \\
    1 &   90 &  2.155$+$01 &  0.000$+$00 &  0.000$+$00 &  0.000$+$00 &  1.414$+$07 &  3.011$+$01 &  0.000$+$00 &   0.0$+$00 \\
    1 &   91 &  2.154$+$01 &  3.598$+$11 &  3.754$-$02 &  1.065$-$02 &  0.000$+$00 &  0.000$+$00 &  4.381$+$04 &   9.5$-$01 \\
    1 &   92 &  2.154$+$01 &  0.000$+$00 &  0.000$+$00 &  0.000$+$00 &  1.498$+$07 &  4.910$+$00 &  0.000$+$00 &   0.0$+$00 \\
    1 &   93 &  2.153$+$01 &  0.000$+$00 &  0.000$+$00 &  0.000$+$00 &  9.335$+$06 &  1.958$+$01 &  0.000$+$00 &   0.0$+$00 \\
    1 &   94 &  2.153$+$01 &  0.000$+$00 &  0.000$+$00 &  0.000$+$00 &  1.171$+$09 &  0.000$+$00 &  0.000$+$00 &   0.0$+$00 \\
    1 &   95 &  2.151$+$01 &  0.000$+$00 &  0.000$+$00 &  0.000$+$00 &  1.606$+$09 &  1.808$-$01 &  0.000$+$00 &   0.0$+$00 \\
    1 &   96 &  2.148$+$01 &  6.229$+$10 &  2.155$-$03 &  6.095$-$04 &  0.000$+$00 &  0.000$+$00 &  1.887$+$04 &   1.3$+$00 \\
    1 &   97 &  2.147$+$01 &  1.437$+$11 &  9.927$-$03 &  2.806$-$03 &  0.000$+$00 &  0.000$+$00 &  4.044$+$04 &   1.3$+$00 \\
    1 &   98 &  2.146$+$01 &  0.000$+$00 &  0.000$+$00 &  0.000$+$00 &  1.386$+$09 &  6.283$+$00 &  0.000$+$00 &   0.0$+$00 \\
    1 &   99 &  2.142$+$01 &  0.000$+$00 &  0.000$+$00 &  0.000$+$00 &  8.041$+$07 &  2.817$+$01 &  0.000$+$00 &   0.0$+$00 \\
    1 &  100 &  2.137$+$01 &  9.936$+$09 &  3.401$-$04 &  9.570$-$05 &  0.000$+$00 &  0.000$+$00 &  6.580$+$04 &   3.2$-$01 \\
    1 &  101 &  2.134$+$01 &  0.000$+$00 &  0.000$+$00 &  0.000$+$00 &  2.070$+$09 &  2.749$+$01 &  0.000$+$00 &   0.0$+$00 \\
    1 &  102 &  2.125$+$01 &  0.000$+$00 &  0.000$+$00 &  0.000$+$00 &  4.964$+$08 &  4.492$+$00 &  0.000$+$00 &   0.0$+$00 \\
\hline  
\end{tabular}
\end{table}

For Sc~XII, $A$-values for all types of transitions, but only among the lowest 27 levels, have been reported by J\"{o}nsson et al. \cite{jon}  and therefore in Table~9 we make comparisons for the E1 and E2 transitions from the lowest 5 to higher excited levels. Generally, for all E1 transitions the agreement between the two calculations is within $\sim$20\%, which is highly satisfactory. However, for three weak transitions, namely 2--11 ($f$ = 1.2$\times$10$^{-4}$), 3--13 ($f$ = 1.2$\times$10$^{-4}$) and 5--9 ($f$ = 4.4$\times$10$^{-5}$), discrepancies are up to a factor of two. As already stated above, accuracies for such weak transitions are often not reliable and hence any of the two calculations can be (in)correct. Similarly, for the comparatively weak E2 transitions the two calculations agree within 20\% for most, but discrepancies are up to a factor of two for four (2--24, 3--24, 3--25, and 5--20), whereas it is factor of four for one, i.e. 5--21 ($f$ = 3.2$\times$10$^{-10}$). Similar comparisons for the M1 and M2 transitions are made in Table~10. There are no  appreciable discrepancies for the M2 transitions (except for 1--25), but for a few M1 the differences are up to two orders of magnitude,   see in particular 2--18 for which our $f$ =  9.7$\times$10$^{-13}$. Such weak transitions (and discrepancies between different calculations) do not affect the modelling, or the subsequent calculations of lifetimes, $\tau$ = 1.0/$\Sigma_{i}$A$_{ji}$,  which includes contributions from all types of transitions, i.e. E1, E2, M1, and M2. This is further confirmed by comparing our results for $\tau$, included in Tables~6--8, for the lowest 27 levels of Sc~XII in Table~11, with those of Hibbert et al. \cite{ah} and J\"{o}nsson et al. \cite{jon}, for which the agreements are within 10\% for most  levels.
 
\clearpage
\setcounter{table}{6} 
\begin{table}
\caption{Transition wavelengths ($\lambda_{ij}$ in \AA), radiative rates (A$_{ji}$ in s$^{-1}$), oscillator strengths (f$_{ij}$, dimensionless), and line     
strengths (S, in atomic units) for electric dipole (E1), and A$_{ji}$ for E2, M1 and M2 transitions in Sc~XII. The last column gives the ratio R  of  the velocity and length forms of A(E1). $a{\pm}b \equiv a{\times}$10$^{{\pm}b}$.}
\small 
\centering
\begin{tabular}{rrrrrrrrrr} \hline        
$i$ & $j$ & $\lambda_{ij}$ & A$^{{\rm E1}}_{ji}$  & f$^{{\rm E1}}_{ij}$ & S$^{{\rm E1}}$ & A$^{{\rm E2}}_{ji}$  & A$^{{\rm M1}}_{ij}$ & A$^{{\rm M2}}$ & R  \\
\hline                                    
    1 &    2 &  3.101$+$01 &  0.000$+$00 &  0.000$+$00 &  0.000$+$00 &  0.000$+$00 &  0.000$+$00 &  2.241$+$04 &   0.0$+$00 \\
    1 &    3 &  3.091$+$01 &  2.365$+$11 &  1.017$-$01 &  1.035$-$02 &  0.000$+$00 &  0.000$+$00 &  0.000$+$00 &   9.4$-$01 \\
    1 &    5 &  3.058$+$01 &  3.359$+$11 &  1.412$-$01 &  1.422$-$02 &  0.000$+$00 &  0.000$+$00 &  0.000$+$00 &   9.4$-$01 \\
    1 &    6 &  2.949$+$01 &  0.000$+$00 &  0.000$+$00 &  0.000$+$00 &  0.000$+$00 &  8.085$+$03 &  0.000$+$00 &   0.0$+$00 \\
    1 &    7 &  2.926$+$01 &  0.000$+$00 &  0.000$+$00 &  0.000$+$00 &  7.312$+$07 &  0.000$+$00 &  0.000$+$00 &   0.0$+$00 \\
    1 &    9 &  2.917$+$01 &  0.000$+$00 &  0.000$+$00 &  0.000$+$00 &  0.000$+$00 &  1.521$+$03 &  0.000$+$00 &   0.0$+$00 \\
    1 &   10 &  2.911$+$01 &  0.000$+$00 &  0.000$+$00 &  0.000$+$00 &  9.485$+$07 &  0.000$+$00 &  0.000$+$00 &   0.0$+$00 \\
    1 &   11 &  2.895$+$01 &  0.000$+$00 &  0.000$+$00 &  0.000$+$00 &  0.000$+$00 &  4.143$+$00 &  0.000$+$00 &   0.0$+$00 \\
    1 &   13 &  2.885$+$01 &  0.000$+$00 &  0.000$+$00 &  0.000$+$00 &  0.000$+$00 &  1.699$+$04 &  0.000$+$00 &   0.0$+$00 \\
    1 &   14 &  2.886$+$01 &  0.000$+$00 &  0.000$+$00 &  0.000$+$00 &  1.167$+$08 &  0.000$+$00 &  0.000$+$00 &   0.0$+$00 \\
    1 &   17 &  2.733$+$01 &  2.044$+$10 &  6.865$-$03 &  6.176$-$04 &  0.000$+$00 &  0.000$+$00 &  0.000$+$00 &   9.8$-$01 \\
    1 &   18 &  2.727$+$01 &  0.000$+$00 &  0.000$+$00 &  0.000$+$00 &  0.000$+$00 &  0.000$+$00 &  5.992$+$05 &   0.0$+$00 \\
    1 &   21 &  2.715$+$01 &  0.000$+$00 &  0.000$+$00 &  0.000$+$00 &  0.000$+$00 &  0.000$+$00 &  6.530$+$04 &   0.0$+$00 \\
    1 &   23 &  2.698$+$01 &  7.276$+$11 &  2.381$-$01 &  2.115$-$02 &  0.000$+$00 &  0.000$+$00 &  0.000$+$00 &   9.8$-$01 \\
    1 &   24 &  2.692$+$01 &  0.000$+$00 &  0.000$+$00 &  0.000$+$00 &  0.000$+$00 &  0.000$+$00 &  6.344$+$04 &   0.0$+$00 \\
    1 &   25 &  2.689$+$01 &  0.000$+$00 &  0.000$+$00 &  0.000$+$00 &  0.000$+$00 &  0.000$+$00 &  1.704$+$03 &   0.0$+$00 \\
    1 &   27 &  2.657$+$01 &  7.503$+$12 &  2.382$+$00 &  2.084$-$01 &  0.000$+$00 &  0.000$+$00 &  0.000$+$00 &   9.8$-$01 \\
    1 &   28 &  2.491$+$01 &  0.000$+$00 &  0.000$+$00 &  0.000$+$00 &  0.000$+$00 &  1.295$+$03 &  0.000$+$00 &   0.0$+$00 \\
    1 &   31 &  2.380$+$01 &  4.824$+$10 &  1.229$-$02 &  9.625$-$04 &  0.000$+$00 &  0.000$+$00 &  0.000$+$00 &   9.9$-$01 \\
    1 &   32 &  2.377$+$01 &  0.000$+$00 &  0.000$+$00 &  0.000$+$00 &  0.000$+$00 &  0.000$+$00 &  8.775$+$04 &   0.0$+$00 \\
    1 &   33 &  2.369$+$01 &  1.114$+$12 &  2.813$-$01 &  2.194$-$02 &  0.000$+$00 &  0.000$+$00 &  0.000$+$00 &   1.0$+$00 \\
    1 &   34 &  2.311$+$01 &  0.000$+$00 &  0.000$+$00 &  0.000$+$00 &  0.000$+$00 &  0.000$+$00 &  6.187$+$03 &   0.0$+$00 \\
    1 &   35 &  2.310$+$01 &  8.909$+$10 &  2.138$-$02 &  1.626$-$03 &  0.000$+$00 &  0.000$+$00 &  0.000$+$00 &   8.2$-$01 \\
    1 &   37 &  2.291$+$01 &  6.570$+$10 &  1.551$-$02 &  1.170$-$03 &  0.000$+$00 &  0.000$+$00 &  0.000$+$00 &   8.3$-$01 \\
    1 &   38 &  2.276$+$01 &  0.000$+$00 &  0.000$+$00 &  0.000$+$00 &  0.000$+$00 &  6.454$+$03 &  0.000$+$00 &   0.0$+$00 \\
    1 &   39 &  2.274$+$01 &  0.000$+$00 &  0.000$+$00 &  0.000$+$00 &  8.428$+$07 &  0.000$+$00 &  0.000$+$00 &   0.0$+$00 \\
    1 &   41 &  2.271$+$01 &  0.000$+$00 &  0.000$+$00 &  0.000$+$00 &  0.000$+$00 &  1.889$+$02 &  0.000$+$00 &   0.0$+$00 \\
    1 &   42 &  2.270$+$01 &  0.000$+$00 &  0.000$+$00 &  0.000$+$00 &  8.206$+$07 &  0.000$+$00 &  0.000$+$00 &   0.0$+$00 \\
    1 &   44 &  2.257$+$01 &  0.000$+$00 &  0.000$+$00 &  0.000$+$00 &  0.000$+$00 &  4.896$+$01 &  0.000$+$00 &   0.0$+$00 \\
    1 &   45 &  2.254$+$01 &  0.000$+$00 &  0.000$+$00 &  0.000$+$00 &  1.516$+$08 &  0.000$+$00 &  0.000$+$00 &   0.0$+$00 \\
    1 &   46 &  2.252$+$01 &  0.000$+$00 &  0.000$+$00 &  0.000$+$00 &  0.000$+$00 &  6.392$+$03 &  0.000$+$00 &   0.0$+$00 \\
    1 &   48 &  2.247$+$01 &  0.000$+$00 &  0.000$+$00 &  0.000$+$00 &  0.000$+$00 &  1.968$+$01 &  0.000$+$00 &   0.0$+$00 \\
    1 &   49 &  2.247$+$01 &  0.000$+$00 &  0.000$+$00 &  0.000$+$00 &  2.945$+$07 &  0.000$+$00 &  0.000$+$00 &   0.0$+$00 \\
    1 &   51 &  2.235$+$01 &  0.000$+$00 &  0.000$+$00 &  0.000$+$00 &  2.082$+$09 &  0.000$+$00 &  0.000$+$00 &   0.0$+$00 \\
    1 &   53 &  2.228$+$01 &  1.660$+$10 &  3.704$-$03 &  2.717$-$04 &  0.000$+$00 &  0.000$+$00 &  0.000$+$00 &   9.5$-$01 \\
    1 &   55 &  2.226$+$01 &  0.000$+$00 &  0.000$+$00 &  0.000$+$00 &  0.000$+$00 &  0.000$+$00 &  3.037$+$05 &   0.0$+$00 \\
    1 &   57 &  2.224$+$01 &  0.000$+$00 &  0.000$+$00 &  0.000$+$00 &  0.000$+$00 &  0.000$+$00 &  5.724$+$04 &   0.0$+$00 \\
\hline  
\end{tabular}
\end{table}

\clearpage
\newpage
\setcounter{table}{6} 
\begin{table}
\caption{... continued.}
\small 
\centering
\begin{tabular}{rrrrrrrrrr} \hline        
$i$ & $j$ & $\lambda_{ij}$ & A$^{{\rm E1}}_{ji}$  & f$^{{\rm E1}}_{ij}$ & S$^{{\rm E1}}$ & A$^{{\rm E2}}_{ji}$  & A$^{{\rm M1}}_{ij}$ & A$^{{\rm M2}}$ & R  \\
\hline 
    1 &   59 &  2.217$+$01 &  1.081$+$12 &  2.389$-$01 &  1.744$-$02 &  0.000$+$00 &  0.000$+$00 &  0.000$+$00 &   9.5$-$01 \\
    1 &   62 &  2.207$+$01 &  0.000$+$00 &  0.000$+$00 &  0.000$+$00 &  0.000$+$00 &  0.000$+$00 &  1.861$+$04 &   0.0$+$00 \\
    1 &   65 &  2.206$+$01 &  0.000$+$00 &  0.000$+$00 &  0.000$+$00 &  0.000$+$00 &  0.000$+$00 &  2.348$+$04 &   0.0$+$00 \\
    1 &   67 &  2.205$+$01 &  0.000$+$00 &  0.000$+$00 &  0.000$+$00 &  2.607$+$07 &  0.000$+$00 &  0.000$+$00 &   0.0$+$00 \\
    1 &   69 &  2.204$+$01 &  0.000$+$00 &  0.000$+$00 &  0.000$+$00 &  0.000$+$00 &  3.594$-$02 &  0.000$+$00 &   0.0$+$00 \\
    1 &   70 &  2.204$+$01 &  0.000$+$00 &  0.000$+$00 &  0.000$+$00 &  3.388$+$07 &  0.000$+$00 &  0.000$+$00 &   0.0$+$00 \\
    1 &   71 &  2.199$+$01 &  2.487$+$12 &  5.406$-$01 &  3.913$-$02 &  0.000$+$00 &  0.000$+$00 &  0.000$+$00 &   9.5$-$01 \\
    1 &   74 &  2.187$+$01 &  0.000$+$00 &  0.000$+$00 &  0.000$+$00 &  7.572$+$07 &  0.000$+$00 &  0.000$+$00 &   0.0$+$00 \\
    1 &   76 &  2.089$+$01 &  0.000$+$00 &  0.000$+$00 &  0.000$+$00 &  0.000$+$00 &  0.000$+$00 &  7.937$+$03 &   0.0$+$00 \\
    1 &   77 &  2.088$+$01 &  5.646$+$10 &  1.108$-$02 &  7.614$-$04 &  0.000$+$00 &  0.000$+$00 &  0.000$+$00 &   7.1$-$01 \\
    1 &   78 &  2.074$+$01 &  0.000$+$00 &  0.000$+$00 &  0.000$+$00 &  0.000$+$00 &  3.896$+$03 &  0.000$+$00 &   0.0$+$00 \\
    1 &   79 &  2.073$+$01 &  0.000$+$00 &  0.000$+$00 &  0.000$+$00 &  2.839$+$07 &  0.000$+$00 &  0.000$+$00 &   0.0$+$00 \\
    1 &   81 &  2.072$+$01 &  0.000$+$00 &  0.000$+$00 &  0.000$+$00 &  0.000$+$00 &  1.077$+$02 &  0.000$+$00 &   0.0$+$00 \\
    1 &   83 &  2.072$+$01 &  0.000$+$00 &  0.000$+$00 &  0.000$+$00 &  2.467$+$07 &  0.000$+$00 &  0.000$+$00 &   0.0$+$00 \\
    1 &   84 &  2.072$+$01 &  4.167$+$10 &  8.049$-$03 &  5.491$-$04 &  0.000$+$00 &  0.000$+$00 &  0.000$+$00 &   7.3$-$01 \\
    1 &   86 &  2.057$+$01 &  0.000$+$00 &  0.000$+$00 &  0.000$+$00 &  0.000$+$00 &  1.494$+$02 &  0.000$+$00 &   0.0$+$00 \\
    1 &   87 &  2.057$+$01 &  0.000$+$00 &  0.000$+$00 &  0.000$+$00 &  0.000$+$00 &  2.694$+$03 &  0.000$+$00 &   0.0$+$00 \\
    1 &   88 &  2.056$+$01 &  0.000$+$00 &  0.000$+$00 &  0.000$+$00 &  3.006$+$07 &  0.000$+$00 &  0.000$+$00 &   0.0$+$00 \\
    1 &   90 &  2.055$+$01 &  1.092$+$10 &  2.073$-$03 &  1.402$-$04 &  0.000$+$00 &  0.000$+$00 &  0.000$+$00 &   9.2$-$01 \\
    1 &   92 &  2.054$+$01 &  0.000$+$00 &  0.000$+$00 &  0.000$+$00 &  0.000$+$00 &  0.000$+$00 &  1.693$+$05 &   0.0$+$00 \\
    1 &   94 &  2.053$+$01 &  0.000$+$00 &  0.000$+$00 &  0.000$+$00 &  0.000$+$00 &  0.000$+$00 &  3.708$+$04 &   0.0$+$00 \\
    1 &   97 &  2.050$+$01 &  9.711$+$11 &  1.835$-$01 &  1.238$-$02 &  0.000$+$00 &  0.000$+$00 &  0.000$+$00 &   9.2$-$01 \\
    1 &   98 &  2.046$+$01 &  0.000$+$00 &  0.000$+$00 &  0.000$+$00 &  0.000$+$00 &  1.006$+$01 &  0.000$+$00 &   0.0$+$00 \\
    1 &  101 &  2.046$+$01 &  0.000$+$00 &  0.000$+$00 &  0.000$+$00 &  7.876$+$07 &  0.000$+$00 &  0.000$+$00 &   0.0$+$00 \\
    1 &  103 &  2.046$+$01 &  0.000$+$00 &  0.000$+$00 &  0.000$+$00 &  1.535$+$08 &  0.000$+$00 &  0.000$+$00 &   0.0$+$00 \\
    1 &  104 &  2.046$+$01 &  0.000$+$00 &  0.000$+$00 &  0.000$+$00 &  0.000$+$00 &  0.000$+$00 &  6.763$-$01 &   0.0$+$00 \\
    1 &  114 &  2.038$+$01 &  0.000$+$00 &  0.000$+$00 &  0.000$+$00 &  0.000$+$00 &  0.000$+$00 &  7.163$+$03 &   0.0$+$00 \\
    1 &  115 &  2.038$+$01 &  0.000$+$00 &  0.000$+$00 &  0.000$+$00 &  0.000$+$00 &  0.000$+$00 &  2.434$+$04 &   0.0$+$00 \\
    1 &  117 &  2.035$+$01 &  1.052$+$12 &  1.960$-$01 &  1.313$-$02 &  0.000$+$00 &  0.000$+$00 &  0.000$+$00 &   9.2$-$01 \\
    1 &  121 &  2.030$+$01 &  0.000$+$00 &  0.000$+$00 &  0.000$+$00 &  1.136$+$08 &  0.000$+$00 &  0.000$+$00 &   0.0$+$00 \\
\hline  
\end{tabular}
\end{table}

For Y~XXX, the only results available in the literature for comparison purposes are the $f$-values of Zhang and Sampson \cite{zs2} for E1 transitions from the ground level and these are compared in Table~12. For a few weak transitions the differences are large, in particular for 1--63 (2s$^2$2p$^6$~$^1$S$_{0}$ -- 2s$^2$2p$^5$4s~$^3$P$^o_{1}$), a spin changing inter-combination transition,  for which our $f$-value is very small ($\sim$10$^{-6}$), and subsequently the discrepancy is of three orders of magnitude. However, such comparisons are very limited and hence cannot be confidently relied upon. In conclusion, on the basis of the (whatever possible)  comparisons have been made for all three ions,  our experience on a wide range of other ions, including F-like \cite{flike1, flike2} and Ne-like \cite{nelike1, nelike2},  and considering that we have included a large CI as well as relativistic effects in generating wavefunctions, we assess the accuracy of our radiative rates to be about 20\%,  for a majority of strong transitions with $f$ $\ge$ 0.1.

\clearpage
\setcounter{table}{7} 
\begin{table}
\caption{Transition wavelengths ($\lambda_{ij}$ in \AA), radiative rates (A$_{ji}$ in s$^{-1}$), oscillator strengths (f$_{ij}$, dimensionless), and line     
strengths (S, in atomic units) for electric dipole (E1), and A$_{ji}$ for E2, M1 and M2 transitions in Y~XXX. The last column gives the ratio R  of  the velocity and length forms of A(E1). $a{\pm}b \equiv a{\times}$10$^{{\pm}b}$.}
\small 
\centering
\begin{tabular}{rrrrrrrrrr} \hline        
$i$ & $j$ & $\lambda_{ij}$ & A$^{{\rm E1}}_{ji}$  & f$^{{\rm E1}}_{ij}$ & S$^{{\rm E1}}$ & A$^{{\rm E2}}_{ji}$  & A$^{{\rm M1}}_{ij}$ & A$^{{\rm M2}}$ & R  \\
\hline                                    
    1 &    2 &  6.212$+$00 &  0.000$+$00 &  0.000$+$00 &  0.000$+$00 &  0.000$+$00 &  0.000$+$00 &  1.133$+$07 &   0.0$+$00 \\
    1 &    3 &  6.201$+$00 &  7.791$+$12 &  1.347$-$01 &  2.750$-$03 &  0.000$+$00 &  0.000$+$00 &  0.000$+$00 &   9.8$-$01 \\
    1 &    4 &  6.036$+$00 &  0.000$+$00 &  0.000$+$00 &  0.000$+$00 &  0.000$+$00 &  2.716$+$07 &  0.000$+$00 &   0.0$+$00 \\
    1 &    5 &  6.028$+$00 &  0.000$+$00 &  0.000$+$00 &  0.000$+$00 &  1.212$+$10 &  0.000$+$00 &  0.000$+$00 &   0.0$+$00 \\
    1 &    7 &  5.976$+$00 &  0.000$+$00 &  0.000$+$00 &  0.000$+$00 &  0.000$+$00 &  2.612$+$06 &  0.000$+$00 &   0.0$+$00 \\
    1 &    9 &  5.971$+$00 &  5.295$+$12 &  8.492$-$02 &  1.669$-$03 &  0.000$+$00 &  0.000$+$00 &  0.000$+$00 &   9.8$-$01 \\
    1 &   10 &  5.962$+$00 &  0.000$+$00 &  0.000$+$00 &  0.000$+$00 &  1.207$+$10 &  0.000$+$00 &  0.000$+$00 &   0.0$+$00 \\
    1 &   12 &  5.810$+$00 &  0.000$+$00 &  0.000$+$00 &  0.000$+$00 &  0.000$+$00 &  9.393$+$05 &  0.000$+$00 &   0.0$+$00 \\
    1 &   14 &  5.772$+$00 &  2.314$+$11 &  3.467$-$03 &  6.588$-$05 &  0.000$+$00 &  0.000$+$00 &  0.000$+$00 &   9.9$-$01 \\
    1 &   16 &  5.758$+$00 &  0.000$+$00 &  0.000$+$00 &  0.000$+$00 &  0.000$+$00 &  0.000$+$00 &  1.542$+$08 &   0.0$+$00 \\
    1 &   18 &  5.756$+$00 &  0.000$+$00 &  0.000$+$00 &  0.000$+$00 &  0.000$+$00 &  1.680$+$07 &  0.000$+$00 &   0.0$+$00 \\
    1 &   19 &  5.752$+$00 &  0.000$+$00 &  0.000$+$00 &  0.000$+$00 &  1.360$+$10 &  0.000$+$00 &  0.000$+$00 &   0.0$+$00 \\
    1 &   20 &  5.748$+$00 &  0.000$+$00 &  0.000$+$00 &  0.000$+$00 &  0.000$+$00 &  0.000$+$00 &  2.417$+$08 &   0.0$+$00 \\
    1 &   23 &  5.704$+$00 &  1.170$+$14 &  1.711$+$00 &  3.213$-$02 &  0.000$+$00 &  0.000$+$00 &  0.000$+$00 &   9.9$-$01 \\
    1 &   24 &  5.560$+$00 &  0.000$+$00 &  0.000$+$00 &  0.000$+$00 &  0.000$+$00 &  0.000$+$00 &  1.046$+$07 &   0.0$+$00 \\
    1 &   25 &  5.549$+$00 &  0.000$+$00 &  0.000$+$00 &  0.000$+$00 &  0.000$+$00 &  0.000$+$00 &  5.876$+$07 &   0.0$+$00 \\
    1 &   27 &  5.527$+$00 &  1.278$+$14 &  1.756$+$00 &  3.195$-$02 &  0.000$+$00 &  0.000$+$00 &  0.000$+$00 &   9.9$-$01 \\
    1 &   28 &  5.437$+$00 &  0.000$+$00 &  0.000$+$00 &  0.000$+$00 &  0.000$+$00 &  2.951$+$06 &  0.000$+$00 &   0.0$+$00 \\
    1 &   31 &  5.297$+$00 &  8.047$+$12 &  1.015$-$01 &  1.771$-$03 &  0.000$+$00 &  0.000$+$00 &  0.000$+$00 &   1.0$+$00 \\
    1 &   32 &  5.255$+$00 &  0.000$+$00 &  0.000$+$00 &  0.000$+$00 &  0.000$+$00 &  0.000$+$00 &  5.146$+$07 &   0.0$+$00 \\
    1 &   33 &  5.249$+$00 &  2.487$+$13 &  3.082$-$01 &  5.326$-$03 &  0.000$+$00 &  0.000$+$00 &  0.000$+$00 &   1.0$+$00 \\
    1 &   34 &  5.096$+$00 &  0.000$+$00 &  0.000$+$00 &  0.000$+$00 &  0.000$+$00 &  3.141$+$05 &  0.000$+$00 &   0.0$+$00 \\
    1 &   35 &  5.094$+$00 &  0.000$+$00 &  0.000$+$00 &  0.000$+$00 &  6.792$+$09 &  0.000$+$00 &  0.000$+$00 &   0.0$+$00 \\
    1 &   37 &  5.070$+$00 &  0.000$+$00 &  0.000$+$00 &  0.000$+$00 &  1.910$+$11 &  0.000$+$00 &  0.000$+$00 &   0.0$+$00 \\
    1 &   38 &  4.572$+$00 &  0.000$+$00 &  0.000$+$00 &  0.000$+$00 &  0.000$+$00 &  0.000$+$00 &  6.736$+$06 &   0.0$+$00 \\
    1 &   39 &  4.570$+$00 &  2.603$+$12 &  2.445$-$02 &  3.679$-$04 &  0.000$+$00 &  0.000$+$00 &  0.000$+$00 &   9.4$-$01 \\
    1 &   40 &  4.532$+$00 &  0.000$+$00 &  0.000$+$00 &  0.000$+$00 &  0.000$+$00 &  1.509$+$07 &  0.000$+$00 &   0.0$+$00 \\
    1 &   41 &  4.530$+$00 &  0.000$+$00 &  0.000$+$00 &  0.000$+$00 &  4.685$+$09 &  0.000$+$00 &  0.000$+$00 &   0.0$+$00 \\
    1 &   43 &  4.518$+$00 &  0.000$+$00 &  0.000$+$00 &  0.000$+$00 &  0.000$+$00 &  3.076$+$06 &  0.000$+$00 &   0.0$+$00 \\
    1 &   44 &  4.515$+$00 &  0.000$+$00 &  0.000$+$00 &  0.000$+$00 &  4.666$+$09 &  0.000$+$00 &  0.000$+$00 &   0.0$+$00 \\
    1 &   47 &  4.473$+$00 &  1.232$+$11 &  1.108$-$03 &  1.632$-$05 &  0.000$+$00 &  0.000$+$00 &  0.000$+$00 &   9.7$-$01 \\
    1 &   49 &  4.470$+$00 &  0.000$+$00 &  0.000$+$00 &  0.000$+$00 &  0.000$+$00 &  0.000$+$00 &  4.720$+$07 &   0.0$+$00 \\
    1 &   51 &  4.468$+$00 &  0.000$+$00 &  0.000$+$00 &  0.000$+$00 &  0.000$+$00 &  0.000$+$00 &  1.694$+$08 &   0.0$+$00 \\
    1 &   53 &  4.457$+$00 &  5.627$+$13 &  5.028$-$01 &  7.379$-$03 &  0.000$+$00 &  0.000$+$00 &  0.000$+$00 &   9.8$-$01 \\
    1 &   54 &  4.445$+$00 &  0.000$+$00 &  0.000$+$00 &  0.000$+$00 &  0.000$+$00 &  6.868$+$04 &  0.000$+$00 &   0.0$+$00 \\
    1 &   56 &  4.444$+$00 &  0.000$+$00 &  0.000$+$00 &  0.000$+$00 &  6.124$+$09 &  0.000$+$00 &  0.000$+$00 &   0.0$+$00 \\
    1 &   59 &  4.442$+$00 &  0.000$+$00 &  0.000$+$00 &  0.000$+$00 &  8.106$+$10 &  0.000$+$00 &  0.000$+$00 &   0.0$+$00 \\
\hline  
\end{tabular}
\end{table}

\clearpage
\newpage
\setcounter{table}{7} 
\begin{table}
\caption{... continued.}
\small 
\centering
\begin{tabular}{rrrrrrrrrr} \hline        
$i$ & $j$ & $\lambda_{ij}$ & A$^{{\rm E1}}_{ji}$  & f$^{{\rm E1}}_{ij}$ & S$^{{\rm E1}}$ & A$^{{\rm E2}}_{ji}$  & A$^{{\rm M1}}_{ij}$ & A$^{{\rm M2}}$ & R  \\
\hline   
    1 &   63 &  4.442$+$00 &  4.538$+$08 &  4.026$-$06 &  5.887$-$08 &  0.000$+$00 &  0.000$+$00 &  0.000$+$00 &   7.4$-$02 \\
    1 &   64 &  4.404$+$00 &  0.000$+$00 &  0.000$+$00 &  0.000$+$00 &  0.000$+$00 &  6.172$+$05 &  0.000$+$00 &   0.0$+$00 \\
    1 &   66 &  4.392$+$00 &  0.000$+$00 &  0.000$+$00 &  0.000$+$00 &  0.000$+$00 &  7.599$+$06 &  0.000$+$00 &   0.0$+$00 \\
    1 &   67 &  4.391$+$00 &  0.000$+$00 &  0.000$+$00 &  0.000$+$00 &  5.185$+$09 &  0.000$+$00 &  0.000$+$00 &   0.0$+$00 \\
    1 &   68 &  4.348$+$00 &  0.000$+$00 &  0.000$+$00 &  0.000$+$00 &  0.000$+$00 &  0.000$+$00 &  4.712$+$06 &   0.0$+$00 \\
    1 &   69 &  4.345$+$00 &  0.000$+$00 &  0.000$+$00 &  0.000$+$00 &  0.000$+$00 &  0.000$+$00 &  3.947$+$07 &   0.0$+$00 \\
    1 &   71 &  4.341$+$00 &  3.645$+$13 &  3.089$-$01 &  4.415$-$03 &  0.000$+$00 &  0.000$+$00 &  0.000$+$00 &   9.8$-$01 \\
    1 &   73 &  4.321$+$00 &  0.000$+$00 &  0.000$+$00 &  0.000$+$00 &  4.337$+$10 &  0.000$+$00 &  0.000$+$00 &   0.0$+$00 \\
    1 &   76 &  4.142$+$00 &  0.000$+$00 &  0.000$+$00 &  0.000$+$00 &  0.000$+$00 &  2.683$+$06 &  0.000$+$00 &   0.0$+$00 \\
    1 &   79 &  4.108$+$00 &  6.095$+$12 &  4.626$-$02 &  6.257$-$04 &  0.000$+$00 &  0.000$+$00 &  0.000$+$00 &   9.7$-$01 \\
    1 &   80 &  4.100$+$00 &  0.000$+$00 &  0.000$+$00 &  0.000$+$00 &  0.000$+$00 &  0.000$+$00 &  4.307$+$07 &   0.0$+$00 \\
    1 &   81 &  4.096$+$00 &  1.361$+$13 &  1.027$-$01 &  1.385$-$03 &  0.000$+$00 &  0.000$+$00 &  0.000$+$00 &   9.6$-$01 \\
    1 &   82 &  4.094$+$00 &  1.530$+$12 &  1.153$-$02 &  1.554$-$04 &  0.000$+$00 &  0.000$+$00 &  0.000$+$00 &   1.1$+$00 \\
    1 &   83 &  4.093$+$00 &  0.000$+$00 &  0.000$+$00 &  0.000$+$00 &  0.000$+$00 &  0.000$+$00 &  1.039$+$07 &   0.0$+$00 \\
    1 &   84 &  4.078$+$00 &  0.000$+$00 &  0.000$+$00 &  0.000$+$00 &  0.000$+$00 &  7.559$+$06 &  0.000$+$00 &   0.0$+$00 \\
    1 &   85 &  4.078$+$00 &  0.000$+$00 &  0.000$+$00 &  0.000$+$00 &  4.121$+$09 &  0.000$+$00 &  0.000$+$00 &   0.0$+$00 \\
    1 &   87 &  4.073$+$00 &  0.000$+$00 &  0.000$+$00 &  0.000$+$00 &  0.000$+$00 &  2.142$+$06 &  0.000$+$00 &   0.0$+$00 \\
    1 &   88 &  4.072$+$00 &  0.000$+$00 &  0.000$+$00 &  0.000$+$00 &  4.314$+$09 &  0.000$+$00 &  0.000$+$00 &   0.0$+$00 \\
    1 &   90 &  4.060$+$00 &  0.000$+$00 &  0.000$+$00 &  0.000$+$00 &  0.000$+$00 &  2.209$+$05 &  0.000$+$00 &   0.0$+$00 \\   
    1 &   91 &  4.060$+$00 &  0.000$+$00 &  0.000$+$00 &  0.000$+$00 &  4.615$+$08 &  0.000$+$00 &  0.000$+$00 &   0.0$+$00 \\
    1 &   94 &  4.055$+$00 &  3.708$+$10 &  2.742$-$04 &  3.660$-$06 &  0.000$+$00 &  0.000$+$00 &  0.000$+$00 &   9.6$-$01 \\
    1 &   96 &  4.054$+$00 &  0.000$+$00 &  0.000$+$00 &  0.000$+$00 &  0.000$+$00 &  0.000$+$00 &  1.795$+$07 &   0.0$+$00 \\
    1 &   98 &  4.053$+$00 &  0.000$+$00 &  0.000$+$00 &  0.000$+$00 &  0.000$+$00 &  0.000$+$00 &  1.017$+$08 &   0.0$+$00 \\
    1 &   99 &  4.054$+$00 &  0.000$+$00 &  0.000$+$00 &  0.000$+$00 &  5.670$+$10 &  0.000$+$00 &  0.000$+$00 &   0.0$+$00 \\
    1 &  101 &  4.049$+$00 &  2.818$+$13 &  2.077$-$01 &  2.769$-$03 &  0.000$+$00 &  0.000$+$00 &  0.000$+$00 &   9.7$-$01 \\
    1 &  102 &  4.044$+$00 &  0.000$+$00 &  0.000$+$00 &  0.000$+$00 &  0.000$+$00 &  0.000$+$00 &  1.852$+$03 &   0.0$+$00 \\
    1 &  105 &  4.043$+$00 &  0.000$+$00 &  0.000$+$00 &  0.000$+$00 &  0.000$+$00 &  2.972$+$04 &  0.000$+$00 &   0.0$+$00 \\
    1 &  107 &  4.043$+$00 &  0.000$+$00 &  0.000$+$00 &  0.000$+$00 &  2.482$+$07 &  0.000$+$00 &  0.000$+$00 &   0.0$+$00 \\
    1 &  110 &  4.042$+$00 &  0.000$+$00 &  0.000$+$00 &  0.000$+$00 &  5.045$+$10 &  0.000$+$00 &  0.000$+$00 &   0.0$+$00 \\
    1 &  120 &  4.035$+$00 &  0.000$+$00 &  0.000$+$00 &  0.000$+$00 &  0.000$+$00 &  0.000$+$00 &  5.833$+$02 &   0.0$+$00 \\
    1 &  123 &  3.990$+$00 &  3.816$+$11 &  2.733$-$03 &  3.590$-$05 &  0.000$+$00 &  0.000$+$00 &  0.000$+$00 &   8.3$-$01 \\
    1 &  124 &  3.975$+$00 &  0.000$+$00 &  0.000$+$00 &  0.000$+$00 &  0.000$+$00 &  3.987$+$05 &  0.000$+$00 &   0.0$+$00 \\
    1 &  126 &  3.970$+$00 &  0.000$+$00 &  0.000$+$00 &  0.000$+$00 &  0.000$+$00 &  4.065$+$06 &  0.000$+$00 &   0.0$+$00 \\
    1 &  127 &  3.970$+$00 &  0.000$+$00 &  0.000$+$00 &  0.000$+$00 &  2.774$+$09 &  0.000$+$00 &  0.000$+$00 &   0.0$+$00 \\
    1 &  128 &  3.952$+$00 &  0.000$+$00 &  0.000$+$00 &  0.000$+$00 &  0.000$+$00 &  0.000$+$00 &  2.570$+$06 &   0.0$+$00 \\
    1 &  129 &  3.951$+$00 &  0.000$+$00 &  0.000$+$00 &  0.000$+$00 &  0.000$+$00 &  0.000$+$00 &  2.334$+$07 &   0.0$+$00 \\
    1 &  131 &  3.950$+$00 &  1.609$+$13 &  1.129$-$01 &  1.468$-$03 &  0.000$+$00 &  0.000$+$00 &  0.000$+$00 &   9.7$-$01 \\
    1 &  133 &  3.941$+$00 &  0.000$+$00 &  0.000$+$00 &  0.000$+$00 &  3.073$+$10 &  0.000$+$00 &  0.000$+$00 &   0.0$+$00 \\
\hline  
\end{tabular}
\end{table}

\clearpage
\newpage
\begin{table}
\caption{Comparison of $A$-values (s$^{-1}$) for some E1 and E2 transitions of  Sc XII. $a{\pm}b \equiv$ $a\times$10$^{{\pm}b}$.}
\small 
\centering
\begin{tabular}{rrrrrrrrr} \hline
\multicolumn{4}{c}{E1} & \multicolumn{4}{c}{E2}  \\  \hline
I & J & GRASP & GRASP2K    &   I & J & GRASP & GRASP2K          \\
\hline
  1 &    3  &   2.365$+$11  &   2.287$+$11   &     1  &    7  &    7.312$+$07  &    7.120$+$07   \\
  1 &    5  &   3.359$+$11  &   3.194$+$11   &     1  &   10  &    9.485$+$07  &    9.295$+$07   \\
  1 &   17  &   2.044$+$10  &   2.134$+$10   &     1  &   14  &    1.167$+$08  &    1.148$+$08   \\
  1 &   23  &   7.276$+$11  &   8.057$+$11   &     2  &   16  &    2.080$+$05  &    2.031$+$05   \\
  1 &   27  &   7.503$+$12  &   6.955$+$12   &     2  &   17  &    1.941$+$05  &    1.896$+$05   \\
  2 &    6  &   1.663$+$09  &   1.656$+$09   &     2  &   18  &    1.449$+$05  &    1.412$+$05   \\
  2 &    7  &   1.523$+$09  &   1.488$+$09   &     2  &   19  &    2.415$+$05  &    2.345$+$05   \\
  2 &    8  &   3.566$+$09  &   3.474$+$09   &     2  &   20  &    1.114$+$05  &    1.050$+$05   \\
  2 &    9  &   4.716$+$08  &   4.470$+$08   &     2  &   21  &    7.334$+$04  &    7.196$+$04   \\
  2 &   10  &   2.548$+$09  &   2.466$+$09   &     2  &   22  &    1.517$+$05  &    1.500$+$05   \\
  2 &   11  &   6.904$+$06  &   3.371$+$06   &     2  &   23  &    3.611$+$04  &    3.392$+$04   \\
  2 &   13  &   7.042$+$08  &   6.743$+$08   &     2  &   24  &    8.309$+$03  &    5.663$+$03   \\
  2 &   14  &   1.582$+$08  &   1.505$+$08   &     2  &   25  &    3.946$+$04  &    3.876$+$04   \\
  3 &    6  &   2.152$+$08  &   2.138$+$08   &     2  &   26  &    1.954$+$04  &    1.841$+$04   \\
  3 &    7  &   1.687$+$09  &   1.641$+$09   &     2  &   27  &    1.531$+$03  &    1.714$+$03   \\
  3 &    9  &   3.017$+$09  &   2.936$+$09   &     3  &   17  &    1.527$+$04  &    1.473$+$04   \\
  3 &   10  &   1.644$+$09  &   1.612$+$09   &     3  &   18  &    6.446$+$04  &    6.357$+$04   \\
  3 &   11  &   5.256$+$07  &   4.151$+$07   &     3  &   20  &    1.297$+$05  &    1.279$+$05   \\
  3 &   12  &   3.794$+$09  &   3.754$+$09   &     3  &   21  &    1.707$+$05  &    1.644$+$05   \\
  3 &   13  &   4.289$+$06  &   6.717$+$06   &     3  &   22  &    1.086$+$05  &    1.011$+$05   \\
  3 &   14  &   1.604$+$08  &   1.471$+$08   &     3  &   23  &    2.610$+$05  &    2.567$+$05   \\
  3 &   15  &   6.657$+$09  &   5.401$+$09   &     3  &   24  &    2.902$+$03  &    2.114$+$03   \\
  4 &    6  &   3.258$+$07  &   3.229$+$07   &     3  &   25  &    3.672$+$01  &    1.934$+$00   \\
  4 &    9  &   6.182$+$07  &   5.744$+$07   &     3  &   26  &    1.303$+$04  &    1.334$+$04   \\
  4 &   11  &   1.884$+$09  &   1.526$+$09   &     3  &   27  &    1.127$+$05  &    9.381$+$04   \\
  4 &   13  &   1.884$+$09  &   1.901$+$09   &     4  &   18  &    8.533$+$03  &    7.901$+$03   \\
  5 &    6  &   2.666$+$07  &   2.651$+$07   &     4  &   21  &    9.462$+$03  &    8.622$+$03   \\
  5 &    7  &   1.877$+$06  &   1.417$+$06   &     4  &   24  &    1.308$+$05  &    1.147$+$05   \\
  5 &    9  &   7.315$+$05  &   1.588$+$06   &     4  &   25  &    1.063$+$05  &    1.161$+$05   \\
  5 &   10  &   1.068$+$08  &   1.011$+$08   &     5  &   17  &    3.377$+$03  &    3.162$+$03   \\
  5 &   11  &   1.529$+$09  &   1.538$+$09   &     5  &   18  &    3.234$+$03  &    2.876$+$03   \\
  5 &   12  &   8.997$+$08  &   8.253$+$08   &     5  &   20  &    8.139$+$00  &    3.477$+$01   \\
  5 &   13  &   1.605$+$09  &   1.499$+$09   &     5  &   21  &    2.163$+$01  &    8.742$+$01   \\
  5 &   14  &   3.472$+$09  &   3.394$+$09   &     5  &   22  &    1.625$+$04  &    1.561$+$04   \\
  5 &   15  &   1.081$+$10  &   9.160$+$09   &     5  &   23  &    2.365$+$04  &    1.994$+$05   \\
	&   &	&			     &     5  &   24  &    1.100$+$05  &    1.186$+$05   \\  
        &   &	&			     &     5  &   25  &    1.267$+$05  &    1.097$+$05   \\  
	&   &	&			     &     5  &   26  &    2.328$+$05  &    2.243$+$05   \\  
	&   &	&			     &     5  &   27  &    3.928$+$05  &    3.671$+$05   \\  
\hline                                                                                          
\end{tabular}                                                                                           
                            
\begin {flushleft}                                                                                      
                    
\begin{tabbing}                                                                                 
aaaaaaaaaaaaaaaaaaaaaaaaaaaaaaaaaaaa\= \kill                                                           
GRASP: Present results with the GRASP code  for 3948   level calculations \\
GRASP2K: Earlier results of  J\"{o}nsson et al. \cite{jon} with the GRASP2K code  \\                    
                                        
\end{tabbing}                                                                                   
\end {flushleft}                           
\end{table}

\clearpage
\newpage
\begin{table}
\caption{Comparison of $A$-values (s$^{-1}$) for some M1 and M2 transitions of  Sc XII. $a{\pm}b \equiv$ $a\times$10$^{{\pm}b}$.}
\small 
\centering
\begin{tabular}{rrrrrrrrr} \hline
\multicolumn{4}{c}{M1} & \multicolumn{4}{c}{M2}  \\  \hline
I & J & GRASP & GRASP2K    &   I & J & GRASP & GRASP2K          \\
\hline
1  &    6  &    8.085$+$03  &    8.399$+$03  &  1  &	2  &	2.241$+$04  &	 2.101$+$04   \\
1  &    9  &    1.521$+$03  &    1.489$+$03  &  1  &   18  &	5.992$+$05  &	 6.093$+$05   \\
1  &   11  &    4.143$+$00  &    5.307$-$01  &  1  &   21  &	6.530$+$04  &	 6.544$+$04   \\
1  &   13  &    1.699$+$04  &    1.720$+$04  &  1  &   24  &	6.344$+$04  &	 5.909$+$04   \\
2  &    3  &    1.336$+$01  &    1.293$+$01  &  1  &   25  &	1.705$+$03  &	 4.340$+$03   \\
2  &    5  &    9.001$+$02  &    0.129$+$02  &  2  &	6  &	3.133$-$01  &	 3.078$-$01   \\
2  &   17  &    6.678$+$00  &    1.452$+$01  &  2  &	8  &	6.485$-$01  &	 6.301$-$01   \\
2  &   18  &    1.267$-$01  &    2.492$+$01  &  2  &	9  &	8.744$-$01  &	 8.605$-$01   \\
2  &   21  &    9.622$-$02  &    1.307$-$01  &  2  &   10  &	1.556$+$00  &	 1.525$+$00   \\
2  &   22  &    2.993$-$01  &    2.444$-$01  &  2  &   11  &	1.852$-$01  &	 1.742$-$01   \\
2  &   23  &    4.006$+$00  &    4.866$+$00  &  2  &   12  &	1.656$+$00  &	 1.691$+$00   \\
2  &   24  &    1.189$-$02  &    7.164$-$01  &  2  &   15  &	1.702$+$01  &	 1.313$+$01   \\
2  &   26  &    3.194$-$01  &    2.629$-$01  &  3  &	8  &	6.833$-$01  &	 6.686$-$01   \\
2  &   27  &    8.176$+$00  &    6.828$+$00  &  3  &   10  &	5.317$-$01  &	 5.253$-$01   \\
3  &    4  &    6.537$+$02  &    6.601$+$02  &  3  &   13  &	3.847$-$01  &	 3.621$-$01   \\
3  &    5  &    1.503$+$02  &    1.505$+$02  &  4  &   14  &	7.596$-$01  &	 7.437$-$01   \\
3  &   16  &    1.093$+$01  &    3.300$+$01  &  5  &   13  &	1.207$+$00  &	 1.199$+$00   \\
3  &   17  &    2.078$+$00  &    1.155$+$01  &  5  &   14  &	8.823$-$01  &	 8.658$-$01   \\
3  &   18  &    3.613$-$01  &    3.387$-$01  &  &  & & &  \\
3  &   23  &    7.572$-$01  &    1.438$+$00  &  &  & & &  \\
3  &   24  &    4.080$-$01  &    2.401$-$01  &  &  & & &  \\
3  &   25  &    2.637$-$01  &    5.492$-$01  &  &  & & &  \\
4  &    5  &    4.290$+$00  &    4.087$+$00  &  &  & & &  \\
4  &   17  &    4.214$-$01  &    4.192$-$01  &  &  & & &  \\
4  &   23  &    1.147$-$03  &    1.255$-$03  &  &  & & &  \\
4  &   27  &    1.108$+$01  &    9.125$+$00  &  &  & & &  \\
5  &   16  &    1.040$+$00  &    1.498$+$00  &  &  & & &  \\
5  &   17  &    1.243$+$00  &    1.247$+$00  &  &  & & &  \\
5  &   18  &    2.251$+$00  &    7.509$-$01  &  &  & & &  \\
5  &   21  &    1.283$-$01  &    3.911$+$00  &  &  & & &  \\
5  &   24  &    2.047$-$01  &    3.525$+$00  &  &  & & &  \\
5  &   27  &    1.828$+$00  &    1.674$+$01  &  &  & & &  \\
\hline                                                                                          
\end{tabular}                                                                                           
                            
\begin {flushleft}                                                                                      
                    
\begin{tabbing}                                                                                 
aaaaaaaaaaaaaaaaaaaaaaaaaaaaaaaaaaaa\= \kill                                                           
GRASP: Present results with the GRASP code for 3948   level calculations \\
GRASP2K: Earlier results of  J\"{o}nsson et al. \cite{jon} with the GRASP2K code  \\                    
                                        
\end{tabbing}                                                                                   
\end {flushleft}                           
\end{table}

\clearpage
\newpage
\begin{table}
\caption{Comparison of lifetimes ($\tau$, s)   for the lowest 27 levels of  Sc XII. $a{\pm}b \equiv$ $a\times$10$^{{\pm}b}$.}
\small 
\centering
\begin{tabular}{rllrrrrrrrr} \hline
Index  &     Configuration        & Level                  & CIV3 & GRASP2K & GRASP  \\
 \hline
    1   &  2s$^2$2p$^6$ 	 &	 $^1$S$  _{0}$     & .......	& ........  &	........  \\
    2   &  2s$^2$2p$^5$3s	 &	 $^3$P$^o_{2}$     & 	    	& 4.760-05  &	4.462-05  \\
    3   &  2s$^2$2p$^5$3s	 &	 $^3$P$^o_{1}$     & 4.22-12	& 4.372-12  &	4.228-12  \\
    4   &  2s$^2$2p$^5$3s	 &	 $^3$P$^o_{0}$     & 	    	& 1.515-03  &	1.530-03  \\
    5   &  2s$^2$2p$^5$3s	 &	 $^1$P$^o_{1}$     & 3.10-12	& 3.131-12  &	2.977-12  \\
    6   &  2s$^2$2p$^5$3p	 &	 $^3$S$  _{1}$     & 5.11-10	& 5.186-10  &	5.161-10  \\
    7   &  2s$^2$2p$^5$3p	 &	 $^3$D$  _{2}$     & 3.15-10	& 3.124-10  &	3.044-10  \\
    8   &  2s$^2$2p$^5$3p	 &	 $^3$D$  _{3}$     & 2.84-10	& 2.879-10  &	2.805-10  \\
    9   &  2s$^2$2p$^5$3p	 &	 $^3$D$  _{1}$     & 3.17-10	& 2.905-10  &	2.816-10  \\
   10   &  2s$^2$2p$^5$3p	 &	 $^3$P$  _{2}$     & 3.27-10	& 2.340-10  &	2.276-10  \\
   11   &  2s$^2$2p$^5$3p	 &	 $^1$P$  _{1}$     & 2.88-10	& 3.216-10  &	3.104-10  \\
   12   &  2s$^2$2p$^5$3p	 &	 $^3$P$  _{0}$     & 2.17-10	& 2.184-10  &	2.131-10  \\
   13   &  2s$^2$2p$^5$3p	 &	 $^3$P$  _{1}$     & 2.42-10	& 2.451-10  &	2.383-10  \\
   14   &  2s$^2$2p$^5$3p	 &	 $^1$D$  _{2}$     & 2.67-10	& 2.627-10  &	2.559-10  \\
   15   &  2s$^2$2p$^5$3p	 &	 $^1$S$  _{0}$     & 6.74-11	& 6.868-11  &	5.725-11  \\
   16   &  2s$^2$2p$^5$3d	 &	 $^3$P$^o_{0}$     & 1.29-10	& 1.311-10  &	1.287-10  \\
   17   &  2s$^2$2p$^5$3d	 &	 $^3$P$^o_{1}$     & 3.23-11	& 3.459-11  &	3.549-11  \\
   18   &  2s$^2$2p$^5$3d	 &	 $^3$P$^o_{2}$     & 1.31-10	& 1.334-10  &	1.303-10  \\
   19   &  2s$^2$2p$^5$3d	 &	 $^3$F$^o_{4}$     & 1.29-10	& 1.312-10  &	1.278-10  \\
   20   &  2s$^2$2p$^5$3d	 &	 $^3$F$^o_{3}$     & 1.20-10	& 1.226-10  &	1.190-10  \\
   21   &  2s$^2$2p$^5$3d	 &	 $^3$F$^o_{2}$     & 1.15-10	& 1.176-10  &	1.142-10  \\
   22   &  2s$^2$2p$^5$3d	 &	 $^3$D$^o_{3}$     & 1.18-10	& 1.207-10  &	1.166-10  \\
   23   &  2s$^2$2p$^5$3d	 &	 $^3$D$^o_{1}$     & 1.20-12	& 1.227-12  &	1.357-12  \\
   24   &  2s$^2$2p$^5$3d	 &	 $^1$D$^o_{2}$     & 1.15-10	& 1.170-10  &	1.131-10  \\
   25   &  2s$^2$2p$^5$3d	 &	 $^3$D$^o_{2}$     & 1.16-10	& 1.190-10  &	1.151-10  \\
   26   &  2s$^2$2p$^5$3d	 &	 $^1$F$^o_{3}$     & 1.21-10	& 1.234-10  &	1.194-10  \\
   27   &  2s$^2$2p$^5$3d	 &	 $^1$P$^o_{1}$     & 1.47-13	& 1.436-13  &	1.331-13  \\
\hline				  				        			
\end{tabular}										             				    
\begin {flushleft}									            			    
\begin{tabbing} 									      	
aaaaaaaaaaaaaaaaaaaaaaaaaaaaaaaaaaaa\= \kill						      	       

CIV3: Earlier results of Hibbert et al.   \cite{ah} with the {\sc civ3} code \\
GRASP2K: Earlier results of J\"{o}nsson et al. \cite{jon} with the {\sc grasp} code \\	
GRASP: Present results  with the {\sc grasp} code for 3948   level calculations \\
	 			      		      	
\end{tabbing}										      	
\end {flushleft}				
\end{table}

\begin{table}
\caption{Comparison of oscillator strengths ($f$-values)  for some transitions of  Y XXX.} 
\small 
\centering
\begin{tabular}{rrrrrrrrr} \hline
I & J & GRASP & DFS   & I & J & GRASP & DFS             \\
 \hline
     1 &   3	&  0.1347 & 0.120  &   1 &  39    &  0.0245 & 0.020  \\
     1 &   9	&  0.0849 & 0.086  &   1 &  47    &  0.0011 & 0.016  \\
     1 &  14	&  0.0035 & 0.010  &   1 &  53    &  0.5028 & 0.443  \\
     1 &  23	&  1.7114 & 0.999  &   1 &  63    &  4.03-6 & 0.003  \\
     1 &  27	&  1.7560 & 2.278  &   1 &  71    &  0.3089 & 0.384  \\
     1 &  31	&  0.1016 & 0.060  &   1 &  79    &  0.0463 & 0.025  \\
     1 &  33	&  0.3082 & 0.304  &   1 &  81    &  0.1027 & 0.103  \\
\hline				  				        			
\end{tabular}	
 \begin {flushleft}
\begin{tabbing} 									      	
aaaaaaaaaaaaaaaaaaaaaaaaaaaaaaaaaaaa\= \kill						      	       
GRASP: Present results with the GRASP code   for 3948   level calculations \\
DFS: Earlier results of Zhang and Sampson \cite{zs2} with the DFS code  \\		 			      		      	
\end{tabbing}										      	
\end {flushleft}
\end{table}

Radiative rates are also presented for four types of transitions, namely E1, E2, M1, and M2. Again, very limited comparisons are possible because of the paucity of other available data.  However, for the majority of strong transitions the accuracy is assessed to be $\sim$20\%,  which is primarily based on comparisons between the length and velocity forms. Any estimates of accuracy for particularly weak transitions with very small $f$-values will be unreliable. The calculated $A$-values have been used to determine lifetimes and have been listed for all levels. No measurements have so far been performed for any level of the three ions concerned, and theoretical results are available for only the lowest 27 levels of Sc~XII, for which there are no  (large) discrepancies. We hope our results listed for a large number of levels/transitions will be useful for the modelling and diagnostics of a variety of plasmas, fusion in particular.

\section{Conclusions}

In this paper energy levels have been reported for three ions, namely F-like Sc~XIII and Ne-like Sc~XII and Y~XXX. For the calculations the {\sc grasp} code has been adopted and CI has been included among a large number of configurations. Additional calculations have also been performed with {\sc fac}, by including even larger CI. This was necessary for accuracy assessments \cite{rev2} because the existing data available for these ions are very limited. Energies have been listed for the lowest 102, 125 and 139 levels of the respective ions, although calculations have been performed for a much larger ranges. This is because beyond these levels is a mixing from other configurations.  However, energies for higher levels can be obtained from the author on request.  On the basis of a variety of comparisons, the listed energies  (in general) are assessed to be accurate to better than 1\% for  most levels.  However, this assessment of accuracy may change if laboratory measurements in future become available for a larger number of levels.


\begin {flushleft}
{\bf Author contributions:} All the work has been done by the author himself.

{\bf Conflict of interests:} There is no conflict of interest with anyone.
\end {flushleft}

\bibliographystyle{mdpi}




\end{document}